\documentclass[aps,pra,twocolumn,superscriptaddress,nofootinbib]{revtex4-2}
\usepackage{graphics}
\usepackage{amsmath}
\usepackage{physics}
\usepackage{graphicx}
\usepackage{color}
\usepackage{amsfonts}
\usepackage{amssymb}
\usepackage{float}

\DeclareMathOperator*{\argmin}{arg\,min}

\begin{document}

\title{Analytical bounds for non-asymptotic asymmetric state discrimination}

\author{Jason L. Pereira} \email{jason.pereira@fi.infn.it}
\affiliation{Department of Computer Science, University of York, York YO10 5GH, UK}
\affiliation{Department of Physics and Astronomy, University of Florence,
via G. Sansone 1, I-50019 Sesto Fiorentino (FI), Italy}

\author{Leonardo Banchi}
\affiliation{Department of Physics and Astronomy, University of Florence,
via G. Sansone 1, I-50019 Sesto Fiorentino (FI), Italy}
\affiliation{ INFN Sezione di Firenze, via G. Sansone 1, I-50019, Sesto Fiorentino (FI), Italy }

\author{Stefano Pirandola}
\affiliation{Department of Computer Science, University of York, York YO10 5GH, UK}

\date{\today}

\begin{abstract}
    Two types of errors can occur when discriminating pairs of quantum states. Asymmetric state discrimination involves minimizing the probability of one type of error, subject to a constraint on the other. We give explicit expressions bounding the set of achievable errors, using the trace norm, the fidelity, and the quantum Chernoff bound. The upper bound is asymptotically tight and the lower bound is exact for pure states. Unlike asymptotic bounds, our bounds give error values instead of exponents, so can give more precise results when applied to finite-copy state discrimination problems.
\end{abstract}

\maketitle

\section{Introduction}\label{section: introduction}

Suppose we want to carry out one-shot discrimination between a pair of quantum states, $\rho_1$ and $\rho_2$. There are two types of errors that we are interested in. The Type I error, which we call $\alpha$, is the probability of identifying the state as $\rho_2$ when it is actually $\rho_1$, whilst the Type II error, which we call $\beta$, is the probability of identifying the state as $\rho_1$ when it is actually $\rho_2$.
There are two basic paradigms of quantum state discrimination: symmetric discrimination, where the aim is to minimize the average measurement error probability, and asymmetric discrimination, where the aim is to minimize the probability of one type of error subject to a constraint on the other.

In the symmetric setting, the optimal error probability is given by the Helstrom theorem \cite{helstrom1969quantum} and depends on the trace distance between $\rho_1$ and $\rho_2$, which can be bounded using the fidelity, via the Fuchs-van der Graaf inequalities~\cite{watrous_theory_2018}, or using the quantum Chernoff bound (QCB)~\cite{audenaert_discriminating_2007}.
For asymmetric discrimination, we have the quantum Neyman-Pearson relation~\cite{audenaert_asymptotic_2008}, which gives the minimum weighted average of the two error types, and thus implicitly lets us find the boundary of the set of achievable errors.

Asymmetric discrimination is needed in situations where one type of error is more undesirable than the other, and is ubiquitous in quantum information. In terms of applications, asymmetric state discrimination is the central model adopted in many protocols of quantum sensing/metrology, including quantum target detection~\cite{karsa_quantum_2020,lloyd_enhanced_2008,tan_quantum_2008,shapiro_quantum_2009,zhuang_optimum_2017,zhuang_entanglement-enhanced_2017,barzanjeh_microwave_2015,guha_gaussian-state_2009,xiong_improve_2017,sanz_quantum_2017,weedbrook_how_2016,ragy_quantifying_2014,wilde_gaussian_2017,de-palma_minimum_2018,lopaeva_experimental_2013,meda_photon-number_2017,zhang_entanglements_2013,zhang_entanglement-enhanced_2015}, quantum reading of memories~\cite{pirandola_quantum_2011,lupo_quantum_2013,spedalieri_cryptographic_2015,pirandola_quantum_2011-1,guha_reading_2013,nair_discriminating_2011,nair_optimal_2011,tej_quantum_2013,bisio_tradeoff_2011,dallarno_experimental_2012,invernizzi_optimal_2011,dallarno_ideal_2012,roga_device-independent_2015}, quantum-enhanced pattern recognition~\cite{banchi_quantum-enhanced_2020}, quantum-enhanced detection of bacterial growth~\cite{spedalieri_detecting_2020}, quantum-enhanced optical super-resolution~\cite{tsang_quantum_2016,lupo_ultimate_2016,nair_far-field_2016,kerviche_fundamental_2017,rehacek_optimal_2017,yang_fisher_2017,lu_quantum-optimal_2018,tang_fault-tolerant_2016,nair_interferometric_2016,yang_far-field_2016,tham_beating_2017,paur_achieving_2016,gatto_monticone_beating_2014,classen_superresolving_2016,tsang_quantum_2009,rozema_scalable_2014}, etc, where it is often more important to avoid false-negatives (fail to spot a target that is present) than false-positives (detect a target when none is present). It is also central in the theory of quantum communications~\cite{pirandola_advances_2020,pirandola2017fundamental}. The decoding capabilities for receivers in quantum communications (e.g. cryptographic) scenarios are directly related to how well they can perform quantum measurements.

Asymmetric state discrimination has largely (though not exclusively~\cite{audenaert_quantum_2012,wang_one-shot_2012}) been studied in the asymptotic regime, where the aim is to find the maximum exponent for the decay rate of one error type, subject to a constraint on the other. This problem has been solved, via the quantum Stein's lemma~\cite{hiai_proper_1991,ogawa_strong_2000,li_second-order_2014,berta_composite_2021} and the quantum Hoeffding bound~\cite{nagaoka_converse_2006,hayashi_error_2007}. These results, however, tell us nothing about the actual values of the two types of errors, only the rates at which they decay. Finite-size analysis is a more realistic treatment for the practical implementation of a number of quantum information protocols.

One-shot expressions for $\alpha$ and $\beta$ can be numerically computed using the hypothesis-testing quantum 
relative entropy, which requires solving a semi-definite program~\cite{wang_one-shot_2012}. However, 
numerical calculations become difficult for high-dimensional systems and scale exponentially with the number of 
copies. Moreover, such calculations are impossible for continuous variable systems, such as Gaussian states. Ref.~\cite{wang_one-shot_2012} also gives a lower bound on the optimal errors, based on the quantum relative entropy (QRE), which is applicable when the QRE is finite and can be computed for Gaussian states~\cite{pirandola2017fundamental}. See the appendices for more information.

In this paper we find upper and lower bounds for $\alpha$ and $\beta$ in terms of the fidelity between $\rho_1$ 
and $\rho_2$. We also find an upper bound based on the QCB. Our bounds have several advantages: i) 
they can be computed efficiently for Gaussian states, using known expressions 
for the fidelity~\cite{banchi2015quantum} and QCB~\cite{pirandola_computable_2008} -- this is crucial 
for applications such as quantum radar~\cite{karsa_quantum_2020,zhuang_entanglement-enhanced_2017,de-palma_minimum_2018}; ii) they can be 
applied without additional computational cost to situations involving 
multi-copy states, since both the fidelity and the QCB can be
expressed in terms of their single-copy values. iii) our QCB bound is asymptotically tight, 
as it saturates the quantum Hoeffding bound, while the fidelity bound is exact for pure states.
Unlike previous results about the asymptotic regime, these bounds allow the receiver operating characteristic to be drawn for both one-shot and multi-shot discrimination between any pair of states. All proofs are in the appendices. A Mathematica notebook containing implementations of the bounds is available as Supplemental Material~\cite{supp}.

\section{Bounds on optimal asymmetric discrimination}\label{section: optimal ROC}

For measurement operators $\Pi_1$ and $\Pi_2$, where $\Pi_2 = \mathbf{I}-\Pi_1$, we can write $\alpha = \Tr[\Pi_2 \rho_1]$ and $\beta = \Tr[\Pi_1 \rho_2]$. 
In terms of an auxiliary parameter $p$, which lies in the range $0\leq p \leq 1$,
the errors are connected via the quantum Neyman-Pearson relation, which states that~\cite{audenaert_asymptotic_2008}
\begin{equation}
	p\alpha + (1-p) \beta \geq p\alpha^* + (1-p) \beta^* = \frac{1-t_p}2,
		\label{eq: neyman pearson}
\end{equation}
where $\{\alpha^*,\beta^*\}$ are a pair of achievable errors that are optimal for a particular value of $p$, in that they minimize $p\alpha+(1-p)\beta$. Such minimization can be solved explicitly and the result written as a function of the trace norm 
\begin{equation}
    t_p = \|(1-p)\rho_2 - p\rho_1\|_1.
		\label{eq: trace norm}
\end{equation}
The optimal errors $\{\alpha^*,\beta^*\}$ are achieved by the POVM
\begin{equation}
    \Pi^*_{1} = \{(1-p)\rho_2 - p\rho_1\}_-,~~~~\Pi^*_{2} = \{(1-p)\rho_2 - p\rho_1\}_+,\label{eq: opt errors}
\end{equation}
where $\{X\}_{\pm}$ is the projector onto the positive/negative eigenspace of $X$ and where we have assumed $((1-p)\rho_2 - p\rho_1)$ is full rank -- the general case is discussed in Appendix~\ref{app: non-differentiable}. 

We want explicit expressions for $\alpha^*$ and $\beta^*$ that let us draw the boundary of the set of achievable error probabilities, rather than the implicit expression given in Eq.~(\ref{eq: neyman pearson}). Such a curve is called the receiver operating characteristic (ROC), and tells us the optimal Type I error for a given Type II error and vice versa.
\begin{figure}[t]
	\centering
	\includegraphics[width=0.4\textwidth]{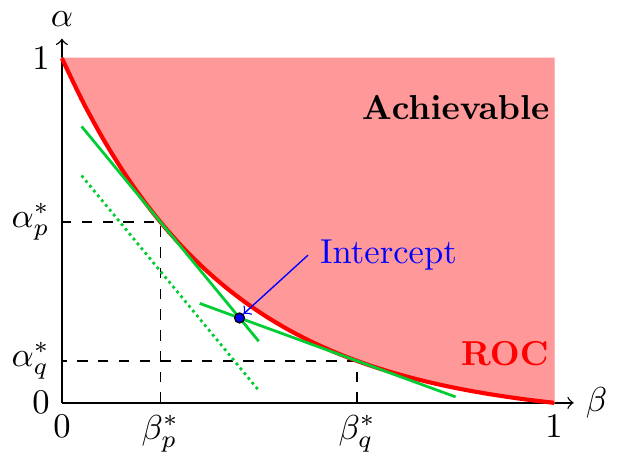}
	\caption{The equation 
		$p \alpha+(1-p)\beta=(1-t_p)/2$ defines a tangent to the ROC curve, which is parametrically defined as
		$(\beta_p^*,\alpha_p^*)$ for varying $p$. 
		By considering two tangents (solid green lines) for different parameters $p$ and $q$, and taking 
		the limit $q\to p$, we can obtain from the intercept one point in the ROC curve, as given in Eq.~\eqref{eq: boundary eq}. 
	}
	\label{fig:roc}
\end{figure}
In Appendix~\ref{app: deriving ROC bounds}, we show that $\alpha^*$ and $\beta^*$ can be obtained from 
Eq.~\eqref{eq: trace norm} and its derivative as 
\begin{equation}
    \alpha^* = \frac{1-t_{p}}{2} - \frac{1-p}{2}\frac{dt_{p}}{dp},
    ~~\beta^* = \frac{1-t_{p}}{2} + \frac{p}{2}\frac{dt_{p}}{dp}.\label{eq: boundary eq}
\end{equation}
A visual proof of the above equation is shown in Fig.~\ref{fig:roc}. Due to the Neyman-Pearson relation
\eqref{eq: neyman pearson} the tangents never pass above the ROC. Accordingly, the ROC curve is convex. For some states, there are values of $p$ at which $t_p$ is not differentiable, but we can still obtain a continuous ROC by replacing the derivative with the subgradient of the trace norm \cite{banchi2020convex}, as per Appendix~\ref{app: non-differentiable}. 

Suppose, instead of an expression for $t_p$, we have an expression that bounds $t_p$ from either above or below. Can we use this to bound $\{\alpha^*,\beta^*\}$? We find that a lower bound on $t_p$ gives an upper bound on the curve defining the boundary of achievable errors and an upper bound on $t_p$ gives a lower bound -- see, for instance, the dotted line in Fig.~\ref{fig:roc}, which has a larger value of $t_p$. We are also guaranteed that if functions $f_1$ and $f_2$ both bound $t_p$ from the same side, and $f_2$ is never tighter than $f_1$, then $f_1$ gives a tighter bound on the set of achievable errors than $f_2$. Finally, as long as the bounding functions are differentiable for all $0<p<1$, it does not matter if $t_p$ is not differentiable for some values of $p$.

\section{Bounds based on the fidelity}\label{section: fidelity}

Let us bound $t_p$ using Fuchs-van der Graaf style inequalities. Quantum fidelity is defined by $F(\rho_1,\rho_2) = \left\| \sqrt{\rho_1}\sqrt{\rho_2} \right\|_1$. Defining
\begin{align}
    &t_p^{(\mathrm{UB},F)} = \sqrt{1-4p(1-p)F(\rho_1,\rho_2)^2},\label{eq: tn UB}\\
    &t_p^{(\mathrm{LB},F)} = 1 - 2\sqrt{p(1-p)}F(\rho_1,\rho_2),\label{eq: tn LB}
\end{align}
we get the bounds $t_p^{(\mathrm{LB},F)} \leq t_p \leq t_p^{(\mathrm{UB},F)}$. If both states are pure, the upper bound is an equality.

Using these bounds, we get the expressions
\begin{align}
    &\alpha^{(\mathrm{LB},F)} = \frac{2(1-p)F^2-1+\sqrt{1-4p(1-p)F^2}}{2\sqrt{1-4p(1-p)F^2}},\label{eq: alpha fid LB param}\\
    &\beta^{(\mathrm{LB},F)} = \frac{2p F^2-1+\sqrt{1-4p(1-p)F^2}}{2\sqrt{1-4p(1-p)F^2}},\label{eq: beta fid LB param}
\end{align}
which provide a lower bound on the boundary of the set of achievable errors, and
\begin{align}
    &\alpha^{(\mathrm{UB},F)} = \frac{F}{2}\sqrt{\frac{1-p}{p}},
    &\beta^{(\mathrm{UB},F)} = \frac{F}{2}\sqrt{\frac{p}{1-p}},\label{eq: fid UB param}
\end{align}
which provide an upper bound on the boundary of the set of achievable errors.

To eliminate $p$, we substitute the expressions for the bounds on $\beta$ into the expressions for the bounds on $\alpha$. The lower bound becomes
\begin{equation}
    \alpha^{(\mathrm{LB},F)} = \beta - 2\beta F^2 + F\left(F-2\sqrt{(1-\beta)\beta(1-F^2)}\right),\label{eq: error fid LB}
\end{equation}
whilst the upper bound becomes $\alpha^{(\mathrm{UB},F)} = \frac{1}{4}F^2\beta^{-1}$. The lower bound meets the axes (of the ROC) at the points $(0,F^2)$ and $(F^2,0)$, and is tight for pure states.

\begin{figure}[tb]
\vspace{+0.1cm}
\centering
\includegraphics[width=0.9\linewidth]{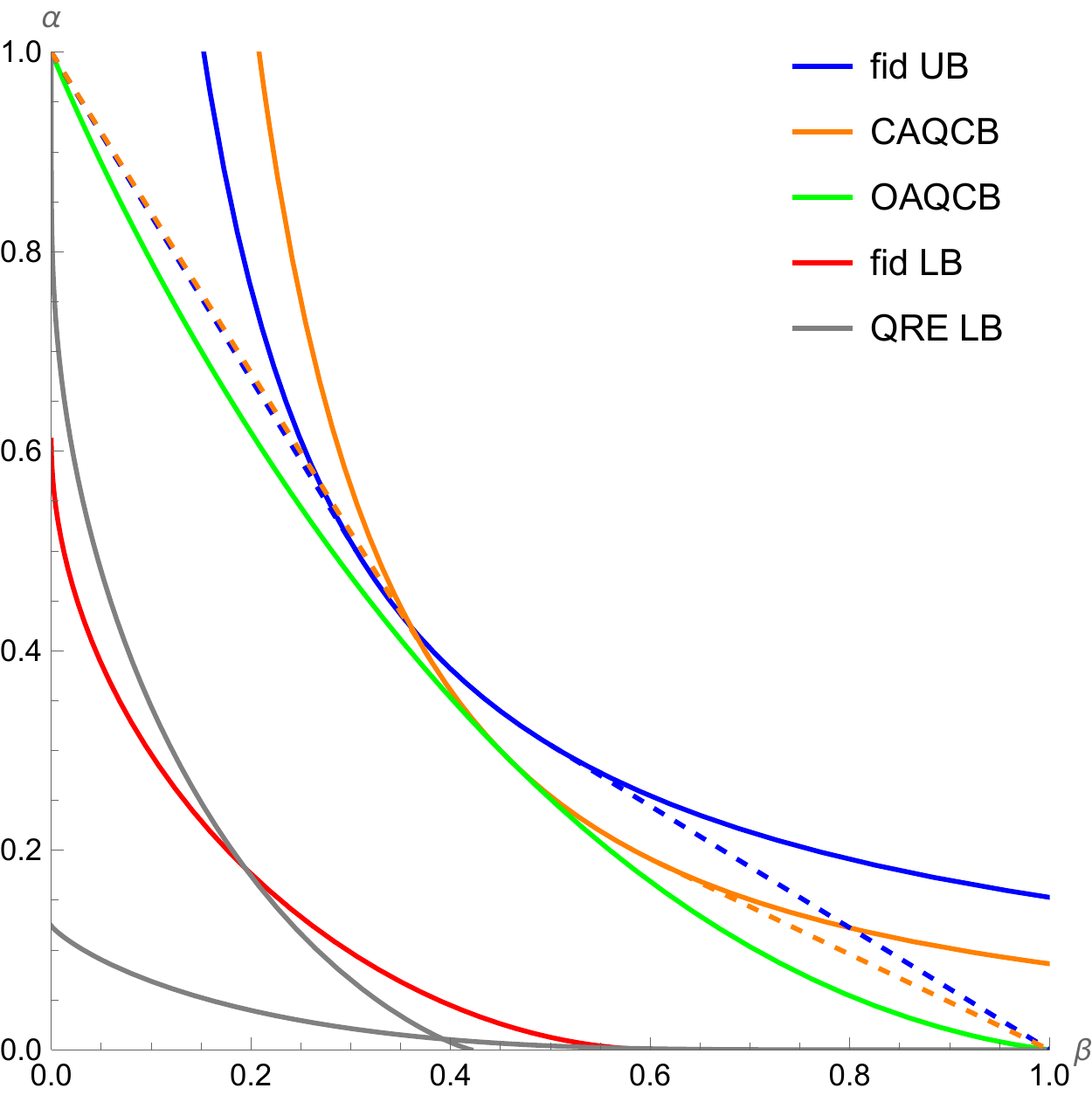}\caption{ROC for discriminating between a pair of states, each the result of transmitting one mode of a two-mode squeezed vacuum, with an average photon number (per mode) of $4$, through a thermal loss channel. $\rho_1$ ($\rho_2$) is obtained using a channel with a transmissivity of $0.7$ ($0.3$) and a thermal number of $0.4$ ($0.6$)~\cite{weedbrook_gaussian_2012}. ``fid UB" and ``fid LB" are the upper and lower bounds based on the fidelity. ``CAQCB" is the upper bound obtained by setting $s_0=s_*$ in Eq.~(\ref{eq: QCB bound piecewise}). For the fidelity upper bound and the CAQCB, the solid lines are the original bounds, whilst the dashed lines are the piecewise modifications. ``QRE LB" is an existing lower bound based on the QRE, from Ref.~\cite{wang_one-shot_2012}.}
\label{fig: bounds}
\end{figure}

The upper bound $\alpha^{(\mathrm{UB},F)} $
diverges to infinity as $p\to 0$ and $\beta^{(\mathrm{UB},F)}$ diverges as $p\to 1$. This is non-physical, because the maximum possible error probability is $1$. Since points $(0,1)$ and $(1,0)$ are achievable and both the ROC and $\alpha^{(\mathrm{UB},F)}$ are convex, we can improve the upper bound using the two tangents to $\alpha^{(\mathrm{UB},F)}$ that pass through points $(0,1)$ and $(1,0)$ (one through each point) -- see the dashed lines in Fig.~\ref{fig: bounds}. We get the tighter piecewise function
\begin{equation}
    \alpha^{(\mathrm{UB},F)} = \begin{cases}
			1-\frac{\beta}{F^2} &\mathrm{~for~} 0\leq \beta \leq \frac{F^2}{2},\\
			\frac{F^2}{4\beta} &\mathrm{~for~} \frac{F^2}{2}\leq \beta\leq \frac12, \\
			(1-\beta)F^2 &\mathrm{~for~} \frac{1}{2}\leq \beta \leq 1.
    \end{cases}\label{eq: fid bound piecewise}
\end{equation}
These bounds are illustrated in Fig.~\ref{fig: bounds}, for a particular pair of states with a fidelity of $\sim 0.782$.

\section{Bounds based on the Quantum Chernoff Bound}\label{section: QCB}

Since the (non-logarithmic) QCB gives a tighter lower bound on the trace distance than the Fuchs-van der Graaf bound, we might suspect it could provide a tighter upper bound on the ROC. Defining
\begin{equation}
    Q_s(\rho_1,\rho_2) = \Tr[\rho_2^s \rho_1^{1-s}],\label{eq: Qs def}
\end{equation}
the QCB, $Q_*$, is given by
\begin{equation}
    Q_* = Q_{s_*}, ~~s_* = \argmin_{0\leq s \leq 1} Q_s.
\end{equation}
In Appendix~\ref{app: QCB bounds}, we show that 
\begin{equation}
    t_p \geq 1 - 2p^{1-s} (1-p)^s Q_s.\label{eq: QCB lower}
\end{equation}
This defines a whole family of bounds. We are particularly interested in two scenarios: fixing $s$ to some set value, $s=s_0$, and setting $s=s_{\mathrm{opt}}$, the value of $s$ that minimizes the right-hand side of Eq.~(\ref{eq: QCB lower}). Note that $s_*$ is a constant, while $s_{\mathrm{opt}}$ is a function of $p$.
For constant $s=s_0$
\begin{align}
    &\alpha^{(\mathrm{UB},s_0)} = \left(\frac{1-p}{p}\right)^{s_0} (1-s_0)Q_{s_0},\label{eq: alpha QCB const}\\
    &\beta^{(\mathrm{UB},s_0)} = \left(\frac{p}{1-p}\right)^{1-s_0} s_0 Q_{s_0}.\label{eq: beta QCB const}
\end{align}
Eliminating $p$, we get
\begin{equation}
    \alpha^{(\mathrm{UB},s_0)} = (1-s_0)Q_{s_0}^{\frac{1}{1-s_0}}\left(\frac{s_0}{\beta}\right)^{\frac{s_0}{1-s_0}}.\label{eq: QCB fixed UB}
\end{equation}
In particular, we might consider setting $s_0=s_*$, so that $Q_{s_0}=Q_*$ (the QCB).

This family of bounds (fixed $s=s_0$) diverges when one of the errors is small (similarly to the fidelity-based upper bound). We can again formulate piecewise bounds, using the tangents to these curves that pass through points $(0,1)$ and $(1,0)$:
\begin{equation}
    \alpha^{(\mathrm{UB},s_0)} = \begin{cases}
    1-\beta Q_{s_0}^{-\frac{1}{s_0}} &0\leq \beta \leq s_0 Q_{s_0}^{\frac{1}{s_0}},\\
    (1-s_0)Q_{s_0}^{\frac{1}{1-s_0}}\left(\frac{s_0}{\beta}\right)^{\frac{s_0}{1-s_0}} & s_0 Q_{s_0}^{\frac{1}{s_0}}\leq \beta\leq s_0,\\
    (1-\beta)Q_{s_0}^{\frac{1}{1-s_0}} &s_0\leq \beta \leq 1.
    \end{cases}\label{eq: QCB bound piecewise}
\end{equation}
We call this family of bounds the constant asymmetric QCBs (CAQCBs).

The bound obtained by setting $s=s_{\mathrm{opt}}$ is
\begin{align}
    &\alpha^{(\mathrm{UB},\mathrm{QCB})} = \exp\left[-p Q_p^{-1}\frac{dQ_p}{dp}\right] (1-p)Q_{p},\label{eq: alpha QCB opt}\\
    &\beta^{(\mathrm{UB},\mathrm{QCB})} = \exp\left[(1-p) Q_p^{-1}\frac{dQ_p}{dp}\right] p Q_{p}.\label{eq: beta QCB opt}
\end{align}
This is the optimal upper bound based on the QCB, so we call it the optimal asymmetric QCB (OAQCB).

Unlike the upper bound based on the fidelity or the CAQCBs, the OAQCB meets the axes at points $(0,Q_1)$ and $(Q_0,0)$ (and $Q_s\leq 1$), so does not require piecewise modification. As demonstrated in Fig.~\ref{fig: bounds}, the OAQCB meets the CAQCB with $s_0$ set to $s_*$ at the point $p=s_*$ (in fact, any CAQCB meets the OAQCB at $p=s_0$).

Explicit expressions for Gaussian states are provided in Appendix~\ref{app: gaussian}, using results from 
Ref.~\cite{banchi2015quantum}.

\section{Multicopy scaling}\label{section: multicopy}

Let us consider how the bounds scale for multi-copy states. The two states we are discriminating between now take the form $\rho_1^{\otimes N}$ and $\rho_2^{\otimes N}$. We are interested in the scaling of the bounds with $N$.

The trace distance between multi-copy states cannot be easily expressed in terms of the single-copy trace distance. On the other hand, both the fidelity and $Q_s$ (as defined in Eq.~(\ref{eq: Qs def})) are simply given by their single-copy values ($F_{(1)}$ and $Q_{s,(1)}$) to the power of $N$. We can write
\begin{equation}
    F_{(N)} = F_{(1)}^N,~~Q_{s,(N)} = Q_{s,(1)}^N.
\end{equation}
This is one major benefit of using bounds based on the fidelity or the QCB rather than the trace norm.

For the bounds based on the fidelity, Eqs.~(\ref{eq: error fid LB}) to (\ref{eq: fid bound piecewise}), we replace $F$ with $F_{(1)}^N$. We find the $N$-copy versions of the CAQCBs, Eqs.~(\ref{eq: QCB fixed UB}) and (\ref{eq: QCB bound piecewise}), in a similar way, by replacing $Q_s$ with $Q_{s,(1)}^N$. 
For the OAQCB we find:
\begin{align}
    &\alpha^{(\mathrm{UB},\mathrm{QCB})}_{(N)} = \frac{\left(\alpha^{(\mathrm{UB},\mathrm{QCB})}_{(1)}\right)^N}{(1-p)^{N-1}},\label{eq: alpha QCB multicopy}\\
    &\beta^{(\mathrm{UB},\mathrm{QCB})}_{(N)} = \frac{\left(\beta^{(\mathrm{UB},\mathrm{QCB})}_{(1)}\right)^N}{p^{N-1}}.\label{eq: beta QCB multicopy}
\end{align}

The quantum Hoeffding bound~\cite{nagaoka_converse_2006,hayashi_error_2007} asymptotically bounds the distinguishability of multi-copy states. It constrains the maximum asymptotic decay rate of $\alpha$, subject to a constraint on the asymptotic decay rate of $\beta$. For a family of discrimination tests on multi-copy states, $\mathcal{T}_N$, with corresponding Type I and II errors, $\{\alpha_N^{\mathcal{T}},\beta_N^{\mathcal{T}}\}$, we define the Type I and II asymptotic decay rates as
\begin{equation}
    \gamma_{\alpha}^{\mathcal{T}} = \lim_{N\to\infty} \frac{ -\ln \left[\alpha_N^{\mathcal{T}}\right]}{N},~~
    \gamma_{\beta}^{\mathcal{T}} = \lim_{N\to\infty} \frac{ -\ln \left[\beta_N^{\mathcal{T}}\right]}{N}.
\end{equation}
The quantum Hoeffding bound then gives the maximum possible value of $\gamma_{\alpha}^{\mathcal{T}}$, subject to a constraint on $\gamma_{\beta}^{\mathcal{T}}$:
\begin{equation}
		b_{\mathrm{max}}(r) =
    \sup_{\{\mathcal{T}\}}\left\{\gamma_{\alpha}^{\mathcal{T}}|\gamma_{\beta}^{\mathcal{T}}\geq r\right\} = 
		\sup_{0\leq s <1} \frac{-sr-\ln[Q_{s,(1)}]}{1-s},
		\label{eq: hoeffding}
\end{equation}
This bound is asymptotic and defines the best achievable scaling with the number of copies, but does not give actual values of $\{\alpha,\beta\}$. It holds for $0<r<S(\rho_1\|\rho_2)$, where $S$ is the quantum relative entropy. Outside this range, the quantum Stein's lemma applies -- see below.

Suppose we have a set of tests that achieve the OAQCB, with $p$ fixed for all $N$. We calculate:
\begin{align}
    &\gamma_{\alpha}^{(\mathrm{UB},\mathrm{QCB})}
    =p Q_{p,(1)}^{-1}\frac{dQ_{p,(1)}}{dp} -\ln \left[Q_{p,(1)}\right],\label{eq: gamma alpha}\\
    &\gamma_{\beta}^{(\mathrm{UB},\mathrm{QCB})}
    =-(1-p) Q_{p,(1)}^{-1}\frac{dQ_{p,(1)}}{dp} -\ln \left[Q_{p,(1)}\right].\label{eq: gamma beta}
\end{align}
Setting $r=\gamma_{\beta}^{(\mathrm{UB},\mathrm{QCB})}$ in Eq.~(\ref{eq: hoeffding}), we find that the maximum is obtained for $s=p$ and hence
\begin{equation}
    b_{\mathrm{max}}(\gamma_{\beta}^{(\mathrm{UB},\mathrm{QCB})})=\gamma_{\alpha}^{(\mathrm{UB},\mathrm{QCB})},
\end{equation}
showing the OAQCB achieves the best possible scaling, according to the quantum Hoeffding bound. Hence, the OAQCB is asymptotically tight, and any tighter upper bound has at most a sub-exponential advantage.

Next we consider the
quantum Stein's lemma~\cite{hiai_proper_1991,ogawa_strong_2000}, which states:
\begin{align*}
		&\sup_{\{\mathcal{T}\}}\left\{\gamma_{\alpha}^{\mathcal{T}}|\gamma_{\beta}^{\mathcal{T}}\geq 0\right\} = S_{21},
		&\sup_{\{\mathcal{T}\}}\left\{\gamma_{\alpha}^{\mathcal{T}}|\gamma_{\beta}^{\mathcal{T}}=S_{12}\right\} = 0,
\end{align*}
where $S_{ij}=S(\rho_i\|\rho_j)$. The maximum exponential rate at which the Type I error decreases with $N$ such that the Type II error does not exponentially increase with $N$ is given by the (single-copy) relative entropy.

We have shown that the OAQCB saturates the quantum Hoeffding bound, however this is only applicable in the range $0<r<S(\rho_1\|\rho_2)$. We can show that it also saturates the quantum Stein's lemma.

Taking the limit of Eqs.~(\ref{eq: gamma alpha}) and (\ref{eq: gamma beta}) as $p\to 0$, we get
\begin{equation}
    \gamma_{\alpha,p\to 0}^{(\mathrm{UB},\mathrm{QCB})} = 0,~~
    \gamma_{\beta,p\to 0}^{(\mathrm{UB},\mathrm{QCB})} = S(\rho_1\|\rho_2).\label{eq: p to 0 lim}
\end{equation}
Taking the limit of Eqs.~(\ref{eq: gamma alpha}) and (\ref{eq: gamma beta}) as $p\to 1$, we get
\begin{equation}
    \gamma_{\alpha,p\to 1}^{(\mathrm{UB},\mathrm{QCB})} = S(\rho_2\|\rho_1),~~
    \gamma_{\beta,p\to 1}^{(\mathrm{UB},\mathrm{QCB})} = 0.\label{eq: p to 1 lim}
\end{equation}
Since $\gamma_{\beta (\alpha)}^{(\mathrm{UB},\mathrm{QCB})}$ is a non-increasing (non-decreasing) function of $p$,
		$0\leq\gamma_{\beta}^{(\mathrm{UB},\mathrm{QCB})}\leq S_{12}$,
		$0\leq\gamma_{\alpha}^{(\mathrm{UB},\mathrm{QCB})}\leq S_{21}$,
for all values of the parameter $p$. Therefore, the OAQCB also saturates the quantum Stein's lemma.

\section{Non-adaptive and adaptive measurement sequences}\label{section: sequences}

Suppose we have one of two pure, $N$-copy states. We know the best possible Type I and II errors are given by Eqs.~(\ref{eq: alpha fid LB param}) and (\ref{eq: beta fid LB param}). We may ask whether these errors can be achieved with a sequence of single-copy measurements, i.e. by carrying out measurements on each copy individually, rather than collective measurements on multiple copies at once.

We consider two types of measurement sequences: non-adaptive sequences, in which the same measurement is carried out on each subsystem, and adaptive sequences, in which the measurement carried out on subsequent subsystems depends on the result of a measurement on a previous subsystem. We assume all of the single-copy measurements are optimal, i.e. have errors given by Eqs.~(\ref{eq: alpha fid LB param}) and (\ref{eq: beta fid LB param}), where $F$ is the single-copy fidelity, $F_{(1)}$.

As an example for non-adaptive sequences, consider a three-copy state. We carry out three measurements - one on each subsystem - and use the results to decide which of two possible states we have. There are three main ways of combining the results to make our decision: decide we have $\rho_1^{\otimes 3}$ only if all three measurements tell us we have $\rho_1$ (case a), majority vote (case b), or decide we have $\rho_2^{\otimes 3}$ only if all three measurements tell us we have $\rho_2$ (case c).
All three cases result in higher errors than the optimum, except at the points given by parameter values $p=0$ (where case a coincides with the optimum) and $p=1$ (where case c coincides with the optimum) -- 
see Appendix~\ref{app: non-adaptive} for the exact expressions and a plot of the ROCs for each case.

Ref.~\cite{acin_multiple_2005} showed that the optimal measurement on a pure, multi-copy state is achievable with an adaptive sequence of single-copy measurements. Ref.~\cite{brandsen_adaptive_2022} made this more general by removing the requirement that each copy be identical.

Suppose we want to discriminate between a pair of pure states, $\rho_1$ and $\rho_2$, that can be written as $\rho_i = \bigotimes_j^N \rho_{i,j}$, where $\rho_{1,j}$ and $\rho_{2,j}$ have the same dimension for all $j$. Both states can be partitioned into $N$ subsystems in the same way (but $\rho_{i,j}$ and $\rho_{i,k}$ need not be identical copies). The optimal measurement can be achieved with an adaptive sequence of $N$ measurements on each subsystem individually. The measurement on the next subsystem only depends on the result of the previous measurement (not the entire sequence of results).

Eqs.~(\ref{eq: alpha fid LB param}) and (\ref{eq: beta fid LB param}) give a simple, alternative way to show this result. Consider the two-subsystem states $\rho_i = \rho_{i,1} \otimes \rho_{i,2}$, where $F_j$ is the fidelity between $\rho_{1,j}$ and $\rho_{2,j}$. Now consider a measurement sequence in which we carry out an optimal measurement (from the curve defined by Eqs.~(\ref{eq: alpha fid LB param}) and (\ref{eq: beta fid LB param})), with parameter $p_0$, on the first subsystem, then carry out another optimal measurement on the second subsystem, with the parameter depending on the previous measurement result. $p^-_1$ ($p^+_1$) is the parameter if the first measurement tells us that the state is $\rho_{1,1}$ ($\rho_{2,1}$). Then, we use only the second measurement result to decide which state we have.

Using Eqs.~(\ref{eq: alpha fid LB param}) and (\ref{eq: beta fid LB param}) to calculate the error probabilities for the measurement sequence and setting
\begin{align}
    p_1^\mp = \frac{1}{2}\left( 1\mp\sqrt{1-4 p_0(1-p_0)F_1^2} \right),\label{eq: p1 expression}
\end{align}
we find that this sequence achieves the optimal errors for discriminating between states with a fidelity of $F_1 F_2$.

If the first subsystem can be partitioned into further subsystems, we can decompose the first measurement into a sequence of individual measurements on subsystems. In general, we choose a parameter, $p_0$, for the measurement on the first subsystem and then measurements on the $i$-th subsystem have a parameter value of
\begin{equation}
    p^\mp_i = \frac{1}{2}\left( 1\mp\sqrt{1-4 p_0(1-p_0) \prod_{j=1}^{i-1} F_{j}^2} \right),
\end{equation}
where the minus (plus) case is used when the $(i-1)$-th measurement indicates that the state is $\rho_1$ ($\rho_2$).

\section{Discussion}\label{section: discussion}

We have presented explicit expressions for the Type I and II errors for discriminating between pairs of quantum states. Unlike asymptotic bounds, these expressions give actual values for the errors, rather than just error exponents. This could be useful for finite-copy scenarios, where the sub-exponential factors could be important. They give ultimate bounds on the performance of receivers, which can be applied to topics such as quantum target detection.

We have given upper and lower bounds on the ROC based on the fidelity, and a family of upper bounds on it based on the QCB (the CAQCBs and the OAQCB). These bounds can be easily calculated analytically for a wide variety of states, including Gaussian states. It is simple to go from the single-copy expressions to multi-copy expressions.

The fidelity lower bound and the OAQCB are of particular interest. Neither are trivial for any parameter value. The fidelity lower bound is exact for pure states, whilst the OAQCB saturates the quantum Hoeffding bound, so is asymptotically tight.

\smallskip
\begin{acknowledgments}
J.~L.~P. and S.~P acknowledge funding from the European Union's Horizon 2020 Research and Innovation Action under grant agreement No. 862644 (FET-OPEN project: Quantum readout techniques and technologies, QUARTET). L.~B. acknowledges funding from the U.S. Department of Energy, Office of Science, National Quantum Information Science Research Centers, Superconducting Quantum Materials and Systems Center (SQMS) under the contract No. DE-AC02-07CH11359. The authors thank Marco Tomamichel for helpful correspondence.
\end{acknowledgments}

%
%\bibliography{bibliography}

\appendix

\section{Derivation of bounds on the receiver operating characteristic}\label{app: deriving ROC bounds}

The quantum Neyman-Pearson relation can be formulated in terms of a parameter $p$, constrained by $0\leq p \leq 1$. We write
\begin{equation}
    \begin{split}
        \mu_p &= p \alpha + (1-p) \beta\\
        &= p \Tr[\Pi_2 \rho_1] + (1-p) \Tr[(\mathbf{I}-\Pi_2) \rho_2]\\
        &= \Tr[p\Pi_2\rho_1 - (1-p)\Pi_2\rho_2]+(1-p)\Tr[\rho_2]\\
        &= 1 - p - \Tr[\Pi_2((1-p)\rho_2 - p\rho_1)].
    \end{split}
\end{equation}
$\mu_p$ can be viewed as the average error probability for a measurement if the source emits state $\rho_1$ with probability $p$ and state $\rho_2$ with probability $1-p$. Minimizing $\mu_p$ over all operators $\Pi_2 \leq \mathbf{I}$, we find that the optimal value, $\mu^*_p$, is achieved by the POVMs
\begin{align}
    &\Pi^*_{1,p} = \{(1-p)\rho_2 - p\rho_1\}_-,\label{eq: povm1}\\
    &\Pi^*_{2,p} = \{(1-p)\rho_2 - p\rho_1\}_+\label{eq: povm2}
\end{align}
(assuming $(1-p)\rho_2 - p\rho_1$ is full rank), and is equal to
\begin{equation}
    \mu^*_p = 1 - p - \Tr[((1-p)\rho_2 - p\rho_1)_+].
\end{equation}
By definition, $\Tr[X] = \Tr[(X)_+] - \Tr[(X)_-]$, so
\begin{equation}
    \begin{split}
        \Tr[(1-p)\rho_2 - p\rho_1] =& \Tr[((1-p)\rho_2 - p\rho_1)_+]\\
        &- \Tr[((1-p)\rho_2 - p\rho_1)_-]\\
        =& 1 - 2p.
    \end{split}\label{eq: trace exp}
\end{equation}
Similarly, $\|X\|_1 = \Tr[(X)_+] + \Tr[(X)_-]$, so
\begin{equation}
    \begin{split}
        t_p =& \|(1-p)\rho_2 - p\rho_1\|_1\\
        =& \Tr[((1-p)\rho_2 - p\rho_1)_+] + \Tr[((1-p)\rho_2 - p\rho_1)_-].
    \end{split}\label{eq: trace norm exp}
\end{equation}
Combining Eqs.~(\ref{eq: trace exp}) and (\ref{eq: trace norm exp}), we get
\begin{equation}
    \Tr[((1-p)\rho_2 - p\rho_1)_+] = \frac{1}{2}(1-2p+t_p),
\end{equation}
and hence
\begin{equation}
    \mu^*_p = \frac{1}{2}(1-t_p).
\end{equation}
Thus, we can express $\mu^*_p$ in terms of $t_p$, the trace norm of $((1-p)\rho_2 - p\rho_1)$.

We therefore know that there exists some achievable pair of errors, $\{\alpha_p,\beta_p\}$, such that
\begin{equation}
    p\alpha_p + (1-p)\beta_p = \frac{1}{2}(1-t_p).
\end{equation}
Further, we know that there exists no pair of errors with a smaller value of $\mu_p$. Therefore, the straight line (in a plot of $\alpha$ versus $\beta$)
\begin{equation}
    \alpha = -\frac{1-p}{p}\beta + \frac{1-t_p}{2p}\label{eq: tangent}
\end{equation}
defines a tangent to the boundary of the set of achievable errors, for any value of $p$ between $0$ and $1$. Any two such tangents will intersect at exactly one point. Two straight lines defined by $y = m_{1(2)}x+c_{1(2)}$ intersect at
\begin{equation}
    x = \frac{c_2-c_1}{m_1-m_2},~~ y = m_1\frac{c_2-c_1}{m_1-m_2} + c_1.
\end{equation}
Let us choose two values of $p$: $p_0$ and $p_0+\delta$. The tangents for these two values of $p$ will intersect at
\begin{align}
    &\alpha_{\mathrm{intersect}} = \frac{1-t_{p_0}}{2} - \frac{1-p_0}{2}\frac{t_{p_0+\delta}-t_{p_0}}{\delta},\\
    &\beta_{\mathrm{intersect}} = \frac{1-t_{p_0}}{2} + \frac{p_0}{2}\frac{t_{p_0+\delta}-t_{p_0}}{\delta}.
\end{align}
By taking the limit as $\delta \to 0$, we get equations for $\alpha^*$ and $\beta^*$ (Eq.~(\ref{eq: boundary eq}) in the main text), the boundary values of the set of achievable errors, in terms of the auxiliary parameter $p$:
\begin{equation*}
    \alpha^* = \frac{1-t_{p}}{2} - \frac{1-p}{2}\frac{dt_{p}}{dp},
    ~~\beta^* = \frac{1-t_{p}}{2} + \frac{p}{2}\frac{dt_{p}}{dp}.
\end{equation*}
Alternatively, we have integrated the expression for the tangents, with regard to $p$.

Suppose that, instead of having an expression for $t_p$, we have an expression that bounds $t_p$ from either above or below. Swapping a lower bound on $t_p$ for $t_p$ in Eq.~(\ref{eq: tangent}) gives the tangent to the boundary of a set of pairs of errors that is contained by the set of achievable errors, and similarly swapping an upper bound on $t_p$ for $t_p$ gives the tangent to the boundary of a set of pairs of errors that contains the set of achievable errors (since it gives a line that is either a tangent to the set of achievable errors or is strictly below it). For the same reason, if functions $f_1$ and $f_2$ both bound $t_p$ from the same side, and $f_2$ is never tighter than $f_1$, then $f_1$ gives a tighter bound on the set of achievable errors than $f_2$.

\section{Non-differentiable trace norm}\label{app: non-differentiable}

Suppose the differential of the trace norm, $t_p$, does not exist for some values of $p$ (note that $t_p$ itself is continuous). How do we modify our expression for the ROC to accommodate this?

First, observe that discontinuities occur only when $(1-p)\rho_2 - p\rho_1$ is not full rank, and that they are due to eigenvalues of $(1-p)\rho_2 - p\rho_1$ ``flipping" from negative to positive or vice versa. This follows from the fact that the eigenvalues of $(1-p)\rho_2 - p\rho_1$ are differentiable functions of $p$, but the trace norm is the sum of their absolute values (and the gradient of the absolute value function has a discontinuity).

Since $(1-p)\rho_2 - p\rho_1$ is not full rank, we must adjust the expressions for the POVM in Eqs.~(\ref{eq: povm1}) and (\ref{eq: povm2}). They become
\begin{align}
    &\Pi^*_{1,p,q} = \{(1-p)\rho_2 - p\rho_1\}_- + q\Pi_0,\\
    &\Pi^*_{2,p,q} = \{(1-p)\rho_2 - p\rho_1\}_+ + (1-q)\Pi_0,
\end{align}
where $0\leq q\leq 1$,
\begin{equation}
	\Pi_0 = \{(1-p)\rho_2 - p\rho_1\}_0,
\end{equation}
and $\{X\}_0$ denotes the kernel of $X$. The remaining equations up to and including Eq.~(\ref{eq: tangent}) still apply (and are independent of the value of $q$). The difference is that the tangent given by Eq.~(\ref{eq: tangent}) now touches the ROC over a line segment rather than at a single point. This is because each value of $q$ gives a different achievable pair of error values $\{\alpha_{p,q},\beta_{p,q}\}$, but all with the same value of $\mu_p$. Eq.~(\ref{eq: tangent}) therefore gives the ROC for this segment (with $p$ and $t_p$ assessed at the discontinuity).

Note that the expression for the ROC given by Eq.~(\ref{eq: boundary eq}) still applies for all values of $p$ for which the differential of $t_p$ exists. Consequently, we can draw the (piecewise) ROC by using Eq.~(\ref{eq: boundary eq}) for all values of $p$ for which the differential of $t_p$ exists and joining up with a straight line segment the two points attained by taking the limit of Eq.~(\ref{eq: boundary eq}) as $p$ approaches the discontinuity from below and from above.

More precisely, calling $x_p = \tr{\Pi_0\rho_1}$ and 
 $y_p = \tr{\Pi_0\rho_2}$, then by definition $(1-p)y_p-px_p=0$. 
 Therefore we may write 
\begin{align}
	\alpha^*_{p,q} &= \frac{1-t_{p}}{2} - \frac{1-p}{2}\frac{dt_{p}}{dp} + qx_p, \\
	~~\beta^*_{p,q} &= \frac{1-t_{p}}{2} + \frac{p}{2}\frac{dt_{p}}{dp} 
	+ (1-q)\frac{p}{1-p}x_p.
\end{align}
Note that $x_p=0$ as long as $\frac{dt_p}{dp}$ is continuous. 
Suppose now that $\frac{dt_p}{dp}$ has a discontinuity at $\tilde p$,
then 	$\alpha^*_{\tilde p\pm}$ (as well as 	$\beta^*_{\tilde p\pm}$)  are different (where $\tilde p\pm$ 
refers to $p$ approaching $\tilde p$ either from right or left). 
Since $x_{\tilde p}\neq 0$ we can vary $q$ to join the
left and right values. Indeed, 
$(\beta_{\tilde p-},\alpha_{\tilde p-})$ and 
$(\beta_{\tilde p+},\alpha_{\tilde p+})$ define two distinct 
points in the $(\beta,\alpha)$ plane, and we can vary $q$ 
to join them with a straight line.

A simpler proof of the same result can be obtained by 
replacing the derivative $\frac{dt}{dp}$ with 
the subgradient $\partial^{q}_p t_p$. 
The subgradient for the trace norm 
was calculated in Ref.~\cite{banchi2020convex}. When $t_p$
is differentiable, then 
$\partial^{q}_p t_p = \frac{dt_p}{dp}$, 
namely the subgradient consists of a single point, the 
derivative. On the other hand, when $\frac{dt_p}{dp}$ is 
discontinuous, the subgradient consists of a region, parameterized 
by $q$, that interpolates between the two discontinuous points 
\begin{equation}
	\partial^{q}_p t_p  = \frac{dt_p}{dp}\Bigg|_{p-} q
	+ \frac{dt_p}{dp}\Bigg|_{p+} (1-q). 
\end{equation}
Writing 
\begin{align}
	\alpha^*_{p,q} &= \frac{1-t_{p}}{2} - \frac{1-p}{2}\partial_p^q t_{p}, \\
	~~\beta^*_{p,q} &= \frac{1-t_{p}}{2} + \frac{p}{2}\partial_p^q t_{p},
\end{align}
we see that we can use $q$ to join the two discontinuous points
in the $(\beta,\alpha)$ plane with a straight line.

\begin{figure}[tb]
\vspace{+0.1cm}
\centering
\includegraphics[width=0.9\linewidth]{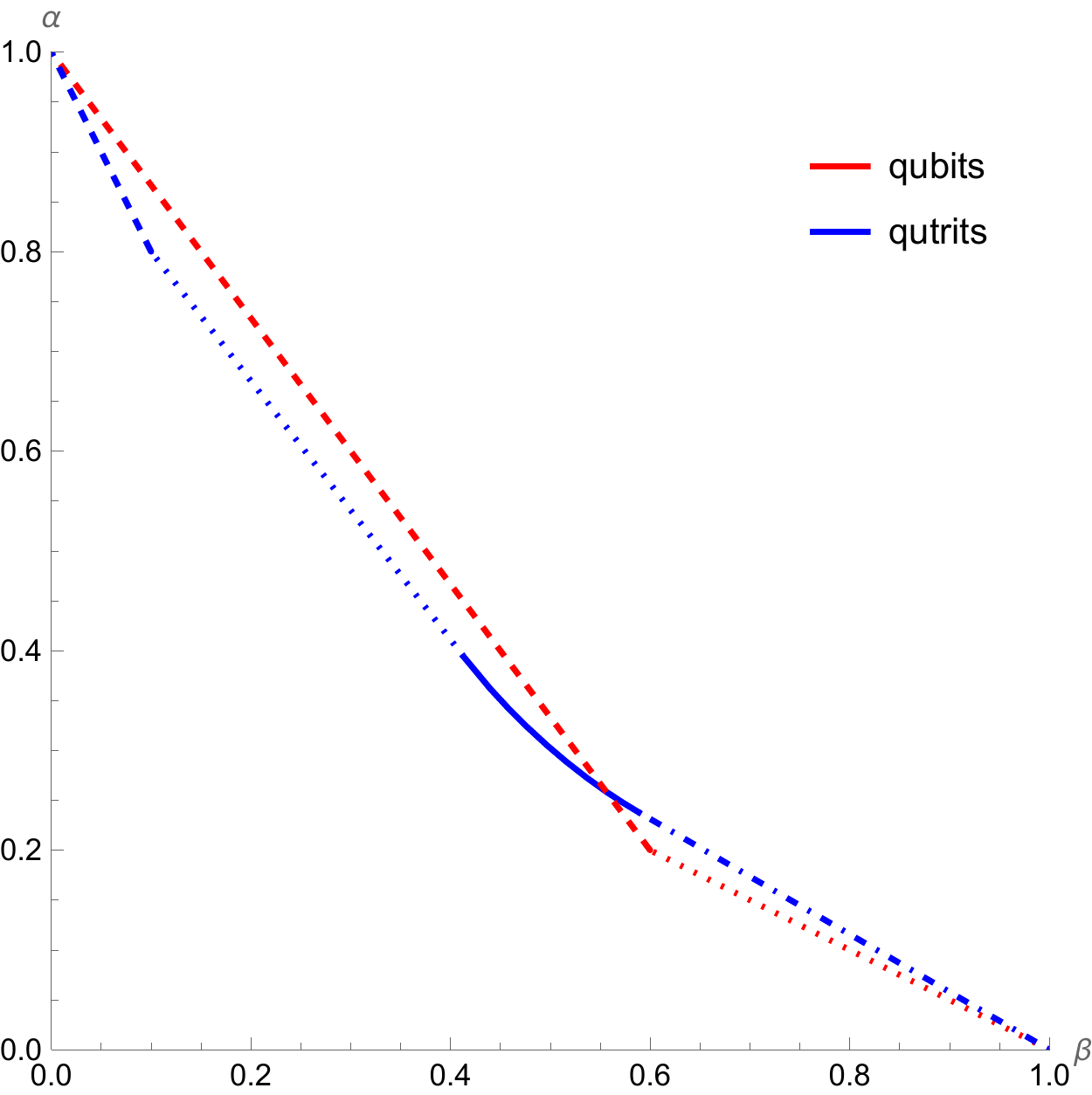}\caption{The ROC for two discrimination problems that both involve a non-differentiable (at points) trace norm, $t_p$. The qubit case involves a pair of qubit states that are diagonalizable in the same basis, whilst the qutrit case involves a pair of qutrit states that are not. The continuous line in the qutrit case is the part of the curve that corresponds to differentiable $t_p$, and is the only part of the ROC that is not a straight line.}
\label{fig: discontinuous}
\end{figure}

Fig.~\ref{fig: discontinuous} gives the ROC for two examples of pairs of states for which $t_p$ is discontinuous. The first is a pair of qubits that are diagonal in the same basis. Specifically, the density matrices of the states are
\begin{equation}
    \rho_1 = \frac{1}{5}\begin{pmatrix}
    4 &0\\
    0 &1
    \end{pmatrix},~~
    \rho_2 = \frac{1}{5}\begin{pmatrix}
    3 &0\\
    0 &2
    \end{pmatrix}.
\end{equation}
The second is a pair of qutrits with density matrices
\begin{equation}
    \rho_1 = \frac{1}{5}\begin{pmatrix}
    3 &0 &0\\
    0 &1 &0\\
    0 &0 &1
    \end{pmatrix},~~
    \rho_2 = \frac{1}{10}\begin{pmatrix}
    6 &1 &1\\
    1 &2 &1\\
    1 &1 &2
    \end{pmatrix}.
\end{equation}

In the qubit case, there are two discontinuities in the gradient of $t_p$ (since there are two value of $p$ for which one of the eigenvalues of $(1-p)\rho_2-p\rho_1$ is equal to $0$). These divide the interval $0\leq p \leq 1$ into three regimes. However, we find that assessing Eq.~(\ref{eq: boundary eq}) in each regime gives a single point with no $p$-dependence in each case. We connect these three points with straight lines to get the curve in Fig.~\ref{fig: discontinuous}.

In the qutrit case, there are three discontinuities, and hence four regimes. However, in this case, in one of the regimes, $\alpha$ and $\beta$ are not constant, but depend on $p$. We therefore have a segment of the ROC that is not a straight line.

\section{Derivation of bounds based on the fidelity}\label{app: fidelity bounds}

We can bound $t_p$ from above and below in terms of the fidelity, by proceeding similarly to the derivations for the $p=\frac{1}{2}$ case in Ref.~\cite{watrous_theory_2018}.

Let $\rho_1$ and $\rho_2$ be a pair of states with fidelity $F$ and let $\left|\rho_1'\right>$ and $\left|\rho_2'\right>$ be purifications of $\rho_1$ and $\rho_2$ that have the same fidelity (these are guaranteed to exist by the definition of fidelity). For any pair of positive semidefinite numbers, $p$ and $q$, and any pair of quantum states, $\left|u\right>$ and $\left|v\right>$, we have the following identity
\begin{equation}
    \left\|p\left|u\middle>\middle<u\right|-q\left|v\middle>\middle<v\right|\right\|_1 = \sqrt{(p+q)^2-4pq|\left<u\middle|v\right>|^2}.
\end{equation}
Therefore, we can write
\begin{equation}
    \begin{split}
        \|(1-p)\left|\rho_2'\middle>\middle<\rho_2'\right|-&p\left|\rho_1'\middle>\middle<\rho_1'\right|\|_1 =\\ &\sqrt{1-4p(1-p)|\left<\rho_2'\middle|\rho_1'\right>|^2}.
    \end{split}
\end{equation}
Then, since the trace norm is monotonic under partial tracing, we have the upper bound
\begin{equation}
    t_p \leq \sqrt{1-4p(1-p)F(\rho_1,\rho_2)^2}.
\end{equation}
If $\rho_1$ and $\rho_2$ are pure, this bound becomes an equality.

For any pair of positive semidefinite operators, $X$ and $Y$, we can write
\begin{equation}
    \|X-Y\|_1\geq\left\|\sqrt{X}-\sqrt{Y}\right\|^2_2.
\end{equation}
Consequently,
\begin{equation}
    \begin{split}
        t_p &\geq \left\|\sqrt{1-p}\sqrt{\rho_2}-\sqrt{p}\sqrt{\rho_1}\right\|^2_2\\
        &\geq\Tr\left[\left(\sqrt{1-p}\sqrt{\rho_2}-\sqrt{p}\sqrt{\rho_1}\right)^2\right]\\
        &\geq (1-p)+p-2\sqrt{p(1-p)}\Tr[\sqrt{\rho_1}\sqrt{\rho_2}]\\
        &\geq 1 - 2\sqrt{p(1-p)}F(\rho_1,\rho_2).
    \end{split}\label{eq: fuchs van der graaf lower}
\end{equation}

We define (Eqs.~(\ref{eq: tn UB}) and (\ref{eq: tn LB}) in the main text)
\begin{align*}
    &t_p^{(\mathrm{UB},F)} = \sqrt{1-4p(1-p)F(\rho_1,\rho_2)^2},\\
    &t_p^{(\mathrm{LB},F)} = 1 - 2\sqrt{p(1-p)}F(\rho_1,\rho_2).
\end{align*}
We can now differentiate both with regard to $p$. We get
\begin{align}
    &\frac{d t_p^{(\mathrm{UB},F)}}{dp}=\frac{2(2p-1)F(\rho_1,\rho_2)^2}{\sqrt{1-4p(1-p)F(\rho_1,\rho_2)^2}},\label{eq: differential of trace norm fid UB}\\
    &\frac{d t_p^{(\mathrm{LB},F)}}{dp}=\frac{(2p-1)F(\rho_1,\rho_2)}{\sqrt{p(1-p)}}.\label{eq: differential of trace norm fid LB}
\end{align}

Substituting Eqs.~(\ref{eq: tn UB}) and (\ref{eq: tn LB}) and Eqs.~(\ref{eq: differential of trace norm fid UB}) and (\ref{eq: differential of trace norm fid LB}) into Eq.~(\ref{eq: boundary eq}), we get upper and lower bounds on the boundary of the set of achievable errors (Eqs.~(\ref{eq: alpha fid LB param}) to (\ref{eq: fid UB param}) in the main text):
\begin{align*}
    &\alpha^{(\mathrm{LB},F)} = \frac{2(1-p)F^2-1+\sqrt{1-4p(1-p)F^2}}{2\sqrt{1-4p(1-p)F^2}},\\
    &\beta^{(\mathrm{LB},F)} = \frac{2p F^2-1+\sqrt{1-4p(1-p)F^2}}{2\sqrt{1-4p(1-p)F^2}},\\
    &\alpha^{(\mathrm{UB},F)} = \frac{F}{2}\sqrt{\frac{1-p}{p}},
    \quad\beta^{(\mathrm{UB},F)} = \frac{F}{2}\sqrt{\frac{p}{1-p}}.
\end{align*}

\section{Derivation of bounds based on the Quantum Chernoff Bound}\label{app: QCB bounds}

From Ref.~\cite{audenaert_discriminating_2007}, we have that, for any pair of positive semidefinite operators, $A$ and $B$, and any $0\leq s\leq 1$,
\begin{equation}
    \Tr[A^s B^{1-s}]\geq \frac{1}{2}\Tr[A+B-|A-B|].
\end{equation}
Substituting $(1-p)\rho_2$ for $A$ and $p\rho_1$ for $B$ and rearranging, we get
\begin{equation}
    p^{1-s} (1-p)^s \Tr[\rho_2^s \rho_1^{1-s}] + \frac{1}{2}\|(1-p)\rho_2-p\rho_1\|_1 \geq \frac{1}{2}.
\end{equation}
Using the definition of $Q_s$, we can therefore write (Eq.~(\ref{eq: QCB lower}) in the main text)
\begin{equation*}
    t_p \geq 1 - 2p^{1-s} (1-p)^s Q_s.
\end{equation*}
If we set $s=s_*$, Eq.~(\ref{eq: QCB lower}) becomes
\begin{equation}
    t_p \geq 1 - 2p^{1-s_*} (1-p)^{s_*} Q_*,
\end{equation}
which we expect to be tighter than the inequality in Eq.~(\ref{eq: fuchs van der graaf lower}) for some values of $p$ (in particular, close to $\frac{1}{2}$).

Note that this is not the tightest lower bound on $t_p$, since $s_*$ minimizes $Q_s$ rather than $p^{1-s} (1-p)^{s} Q_s$. The optimal value of $s$ (achieving the tightest bound) is therefore not a constant, but is rather a function of $p$. We call this value $s_{\mathrm{opt}}$, and define
\begin{equation}
    Q_{\mathrm{opt}} = Q_{s_{\mathrm{opt}}}, ~~s_{\mathrm{opt}} = \argmin_{0\leq s \leq 1} p^{1-s}(1-p)^s Q_s.
\end{equation}
By differentiation, $s_{\mathrm{opt}}$ satisfies
\begin{equation}
    \ln \left[\frac{1-p}{p}\right]Q_{\mathrm{opt}} + \left.\frac{dQ_s}{ds}\right|_{s=s_{\mathrm{opt}}}=0.\label{eq: sopt cond}
\end{equation}
If we have an analytical expression for $Q_s$ in terms of $s$, we can analytically calculate $s_{\mathrm{opt}}(p)$ (although we will later show that finding $s_{\mathrm{opt}}(p)$ is not necessary).

Defining the family of lower bounds on $t_p$ as
\begin{equation}
    t_p^{(\mathrm{LB},s)} = 1 - 2p^{1-s} (1-p)^s Q_s,\label{eq: tn qcb}
\end{equation}
and treating $s$ as a function of $p$, we differentiate to get
\begin{equation}
    \begin{split}
        \frac{dt_p^{(\mathrm{LB},s)}}{dp} =& \frac{\partial t_p^{(\mathrm{LB},s)}}{\partial p} + \frac{\partial t_p^{(\mathrm{LB},s)}}{\partial s}\frac{ds}{dp}\\
        =& -2\frac{(1-p)^{s-1}}{p^s}\bigg((1-p-s)Q_s\\
        &+ p(1-p)\frac{ds}{dp}\left( \ln \left[\frac{1-p}{p}\right]Q_s + \frac{dQ_s}{ds} \right)\bigg).
    \end{split}\label{eq: tn qcb diff}
\end{equation}

We are interested in two scenarios in particular: fixing $s$ to some set value, $s=s_0$, and setting $s=s_{\mathrm{opt}}$. In the former case, $\frac{ds}{dp}=0$, and in the latter case $\frac{\partial t_p^{(\mathrm{LB},s)}}{\partial s}=0$. In both cases, Eq.~(\ref{eq: tn qcb diff}) reduces to
\begin{equation}
    \frac{dt_p^{(\mathrm{LB},s')}}{dp} = -2Q_{s'}\left(\frac{1-p}{p}\right)^{s'}\frac{1-p-s'}{1-p},
\end{equation}
where $s'$ stands in for either $s_0$ or $s_{\mathrm{opt}}$.

Substituting Eqs.~(\ref{eq: tn qcb}) and (\ref{eq: tn qcb diff}) into Eq.~(\ref{eq: boundary eq}), we get
\begin{align}
    &\begin{split}
        \alpha^{(\mathrm{UB},s)} =& \left(\frac{1-p}{p}\right)^{s} \bigg[ (1-s)Q_s +\\
        &p(1-p)\frac{ds}{dp}\left(\ln\left[\frac{1-p}{p}\right]Q_s+\frac{dQ_s}{ds}\right) \bigg],
    \end{split}\label{eq: alpha UB QCB}\\
    &\begin{split}
        \beta^{(\mathrm{UB},s)} =& \left(\frac{p}{1-p}\right)^{1-s} \bigg[ s Q_s +\\
        &p(1-p)\frac{ds}{dp}\left(\ln\left[\frac{p}{1-p}\right]Q_s-\frac{dQ_s}{ds}\right) \bigg].
    \end{split}\label{eq: beta UB QCB}
\end{align}
If we again set $s=s_0$ or $s=s_{\mathrm{opt}}$, we get
\begin{align}
    &\alpha^{(\mathrm{UB},s')} = \left(\frac{1-p}{p}\right)^{s'} (1-s')Q_{s'},\label{eq: alpha QCB gen}\\
    &\beta^{(\mathrm{UB},s')} = \left(\frac{p}{1-p}\right)^{1-s'} s' Q_{s'},\label{eq: beta QCB gen}
\end{align}
where $s'$ stands in for either $s_0$ or $s_{\mathrm{opt}}$. The $s=s_0$ case gives Eqs.~(\ref{eq: alpha QCB const}) and (\ref{eq: beta QCB const}) from the main text.

For any value of $s$, there exists some value of $p$ for which $s_{\mathrm{opt}}$ is given by that $s$ value. In other words, we can validly write $p(s_{\mathrm{opt}})$ instead of $s_{\mathrm{opt}}(p)$. This follows from the fact that $\frac{p}{1-p}$ can take any positive value, so we can always choose $p$ such that Eq.~(\ref{eq: sopt cond}) is satisfied, and the convexity of $p^{1-s}(1-p)^s Q_s$ (this can be seen from the decompositions in Eqs.~(\ref{eq: Qs discrete decomp}) and (\ref{eq: Qs continuous decomp})), which means that the point at which Eq.~(\ref{eq: sopt cond}) is satisfied is a minimum.

Using Eq.~(\ref{eq: sopt cond}), we can write
\begin{equation}
    \frac{p}{1-p}=\exp\left[Q_{\mathrm{opt}}^{-1}\left.\frac{dQ_s}{ds}\right|_{s=s_{\mathrm{opt}}}\right],
\end{equation}
and thus can rewrite Eqs.~(\ref{eq: alpha QCB gen}) and (\ref{eq: beta QCB gen}), replacing $p$ as our auxiliary parameter with $s_{\mathrm{opt}}$.
\begin{align}
    &\alpha^{(\mathrm{UB},\mathrm{QCB})} = \left.\left(\exp\left[-s Q_{s}^{-1}\frac{dQ_s}{ds}\right] (1-s)Q_{s}\right)\right|_{s=s_{\mathrm{opt}}},\\
    &\beta^{(\mathrm{UB},\mathrm{QCB})} = \left.\left(\exp\left[(1-s) Q_s^{-1}\frac{dQ_s}{ds}\right] s Q_{s}\right)\right|_{s=s_{\mathrm{opt}}}.
\end{align}
For consistency with the other equations, we redefine $p$ as $s_{\mathrm{opt}}$ and write (Eqs.~(\ref{eq: alpha QCB opt}) and (\ref{eq: beta QCB opt}) in the main text)
\begin{align*}
    &\alpha^{(\mathrm{UB},\mathrm{QCB})} = \exp\left[-p Q_p^{-1}\frac{dQ_p}{dp}\right] (1-p)Q_{p},\\
    &\beta^{(\mathrm{UB},\mathrm{QCB})} = \exp\left[(1-p) Q_p^{-1}\frac{dQ_p}{dp}\right] p Q_{p}.
\end{align*}

\section{Connection between the OAQCB and quantum relative entropy}\label{app: OAQCB and QRE}

Ref.~\cite{audenaert_discriminating_2007} points out a connection between the QCB and the quantum relative entropy (QRE), namely that for $s_{*}$ (the value that minimizes $Q_s(\rho_1,\rho_2)$), the following holds:
\begin{align}
    &S(\tau_{s_*}\|\rho_1) = S(\tau_{s_*}\|\rho_2),\\
    &\tau_{s} = \frac{\rho_2^s \rho_1^{1-s}}{\Tr[\rho_2^s \rho_1^{1-s}]}
    = \frac{\rho_2^s \rho_1^{1-s}}{Q_s},
\end{align}
where $S(A\|B)$, the QRE between $A$ and $B$, is defined (in nats) by
\begin{equation}
    S(A\|B) = \Tr[A\ln A - A\ln B].
\end{equation}
Note that $\tau_s$ is not, in general, a valid quantum state.

In the discrete variable case, we can decompose $Q_s$ as
\begin{equation}
    Q_s = \sum_i c_i \lambda_i^s \mu_i^{1-s},\label{eq: Qs discrete decomp}
\end{equation}
where $c_i$, $\lambda_i$, and $\mu_i$ are all $\geq 0$~\cite{audenaert_discriminating_2007}. The $\lambda_i$ correspond to eigenvalues of $\rho_2$, the $\mu_i$ correspond to eigenvalues of $\rho_1$, and the $c_i$ correspond to squared overlaps between the eigenvectors of $\rho_1$ and $\rho_2$. We could have written this expression with two separate indices for the $\lambda$ values and the $\mu$ values (and a nested sum over both), but we have chosen to combine them into the single index $i$. Similarly, in the continuous variable case, we can decompose $Q_s$ as
\begin{equation}
    Q_s = \int c_x \lambda_x^s \mu_x^{1-s} dx,\label{eq: Qs continuous decomp}
\end{equation}
where $c_x$, $\lambda_x$, and $\mu_x$ are all positive semidefinite functions of $x$, and where the bounds of the integral may be finite or may be infinite.

By differentiating Eq.~(\ref{eq: Qs discrete decomp}) (Eq.~(\ref{eq: Qs continuous decomp}) in the continuous variable case) with regard to $s$, we get
\begin{equation}
    \frac{dQ_s}{ds} = \sum_i c_i \lambda_i^s \mu_i^{1-s} (\ln[\lambda_i]-\ln[\mu_i])\label{eq: Qs diff}
\end{equation}
(with a similar result in the continuous variable case). We can therefore write
\begin{equation}
    S(\tau_{p}\|\rho_1) - S(\tau_{p}\|\rho_2) = Q_p^{-1}\frac{dQ_p}{dp}.
\end{equation}
For instance, if we want the ratio between $\alpha^{(\mathrm{UB},\mathrm{QCB})}$ and $\beta^{(\mathrm{UB},\mathrm{QCB})}$, for some parameter value $p$, we can express it in terms of the QRE as
\begin{equation}
    \frac{\alpha^{(\mathrm{UB},\mathrm{QCB})}}{\beta^{(\mathrm{UB},\mathrm{QCB})}} = \exp\left[ S(\tau_{p}\|\rho_1) - S(\tau_{p}\|\rho_2) \right]\frac{1-p}{p}.
\end{equation}

Taking the limit of Eq.~(\ref{eq: Qs diff}) as $s\to 0$, we get
\begin{equation}
    \left.\frac{dQ_s}{ds}\right|_{s=0} = \sum_i c_i \mu_i (\ln[\lambda_i]-\ln[\mu_i])=-S(\rho_1\|\rho_2),\label{eq: Qs0}
\end{equation}
and taking the limit as $s\to 1$, we get
\begin{equation}
    \left.\frac{dQ_s}{ds}\right|_{s=1} = \sum_i c_i \lambda_i (\ln[\lambda_i]-\ln[\mu_i])=S(\rho_2\|\rho_1).\label{eq: Qs1}
\end{equation}
For some states, the QRE $S(\rho_1\|\rho_2)$ ($S(\rho_2\|\rho_1)$) can diverge. This corresponds to some of the $\lambda_i$ ($\mu_i$) equaling $0$. Note that this only occurs for extremal values of $s$, since we set $0\ln[0]=0$. These are also the states for which $Q_0$, $Q_1$, or both are not equal to $1$.

\section{Gaussian states}\label{app: gaussian}

The $Q_s$ can be computed for Gaussian states using the formalism developed in 
Ref.~\cite{banchi2015quantum}. There it was shown that a Gaussian state $\hat\rho$ can be expressed as 
\begin{equation}
	\hat \rho = \frac{e^{-\frac12 (\hat Q-u) G (\hat Q-u)}}{Z(G)}\equiv\hat\rho(G,u), \label{eq: rho exp}
\end{equation}
where $\hat Q$ are the quadrature operators, $u$ is the first moment vector, with components 
$u_i =\Tr[\hat \rho Q_i]$, 
the matrix $G$ is obtained from the covariance matrix, defined by $V_{ij} =
\Tr[\hat \rho \{\hat Q_i -u_i,\hat Q_j-u_j\}]/2$
where $\{X,Y\}=XY+YX$, as 
\begin{align}
	V	\equiv V(G) &= \frac12\frac{e^{i\Omega G}+\openone}{e^{i\Omega G}-\openone}i\Omega, &
	e^{i\Omega G} = \frac{W-\openone}{W+\openone},
	\label{eq: V G}
\end{align}
where $W=-2V i\Omega$
and 
\begin{align}
	Z(G) &= \det[(e^{i\Omega G/2}-e^{-i\Omega G/2})i\Omega]^{-1/2} \\ & =
	\sqrt{\det(V+i\Omega/2)}. \nonumber
\end{align}
Thanks to the above definitions, the operator $\hat \rho^s$ is proportional to a Gaussian state 
with a rescaled matrix $G$, namely 
\begin{equation}
	\hat\rho(G,u)^s = \hat \rho(sG,u) \frac{Z(sG)}{Z(G)^s}.
\end{equation}
The final ingredient to compute the $Q_s$ is the following formula, 
\begin{equation}
	\Tr[\hat\rho(G_1,u_1)\hat\rho(G_2,u_2)] = \frac{e^{-\delta^T (V(G_1)+V(G_2))^{-1} \delta/2}}{\sqrt{\det[V(G_1)+V(G_2)]}},
\end{equation}
where $\delta=u_1-u_2$. Suppose now we have two Gaussian states $\hat \rho_1$ and $\hat \rho_2$ with first moments $u_i$ and covariance matrices $V_i$. Let $G_i$ be the corresponding matrices in Eq.~\eqref{eq: rho exp}. Then 
by mixing the above formulas we get (see also \cite{banchi2015quantum,seshadreesan2018renyi})
\begin{align}
	Q_s &= \Tr[\hat\rho_1^s\hat\rho_2^{1-s}] \\ \nonumber &= \frac{Z(sG_1)}{Z(G_1)^s}
	\frac{Z[(1-s)G_2]}{Z(G_2)^{1-s}} 	\Tr[\hat\rho(sG_1,u_1)\hat\rho[(1-s)G_2,u_2]] 
		\\& = \frac{Z(sG_1)}{Z(G_1)^s} \frac{Z[(1-s)G_2]}{Z(G_2)^{1-s}} 
\frac{e^{-\delta^T (V_1(s)+V_2(1-s))^{-1} \delta/2}}{\sqrt{\det[V_1(s)+V_2(1-s)]}},
\nonumber
\end{align}
where using \eqref{eq: V G} we have defined the matrices 
\begin{align}
	V_j(s) &= V(s G_j) = 
	\frac12\frac{e^{i\Omega sG_j}+\openone}{e^{i\Omega sG_j}-\openone}i\Omega \\&= 
 \frac12\frac{\left(\frac{-2V_j i\Omega-\openone}{-2V_ji\Omega+\openone}\right)^s+\openone}{
	 \left(\frac{-2V_j i\Omega-\openone}{-2V_ji\Omega+\openone}\right)^s
-\openone}i\Omega.
\end{align}
This can be simplified by defining 
\begin{equation}
	W_j(s) = -2 V_j(s) i\Omega = 
 \frac{\left(W_j+\openone\right)^s+\left(W_j-\openone\right)^s}{
	 \left(W_j+\openone\right)^s -\left(W_j-\openone\right)^s}.
\end{equation}
Note also that $Z(sG_j) = \sqrt{\det[V_j(s)+i\Omega/2]}$, so we get 
\begin{align}
	\nonumber
	Q_s = &\sqrt{\frac{\det[V_1(s)+i\Omega/2]}{\det[V_1+i\Omega/2]^s}
	\frac{\det[V_2(1-s)+i\Omega/2]}{\det[V_2+i\Omega/2]^{1-s}}} \times \\&\times
\frac{e^{-\delta^T (V_1(s)+V_2(1-s))^{-1} \delta/2}}{\sqrt{\det[V_1(s)+V_2(1-s)]}}.
\end{align}
Taking the logarithm and using the identity $\log(\det A) = \Tr(\log A)$
\begin{align}
    \begin{split}
        -2\log Q_s &= \delta^T (V_1(s)+V_2(1-s))^{-1}\delta  
					 \\& + \Tr\log[V_1(s)+V_2(1-s)]  \\&
	    +2 [s \log Z(G_1) + (1-s)\log Z(G_2)]  \\& - 
	    \Tr\log(V_1(s)+i\Omega/2)  \\& - \Tr\log(V_2(1-s)+i\Omega/2).
    \end{split}
\end{align}
To compute derivatives we note that
\begin{equation}
	W_j'(s) = \frac{2 (W_j-\openone)^s (W_j+\openone)^s }{\left((W_j-\openone)^s-(W_j+\openone)^s\right)^2}\log\frac{W_j-\openone}{W_j+\openone}
\end{equation}
and $V_j'(s) = -W_j'(s)i\Omega/2$. Moreover, using properties of Fr\'echet derivatives $\Tr[f(X(s))]= 
\Tr [f'(X(s)) X'(s)]$, where $f$ is a function and $X$ a matrix, so 
\begin{align}
	g_j(s) &= -\partial_s \Tr\log(V_j(s)+i\Omega/2)
			\\ &= \Tr\left[
\frac{(W_j+\openone)^s %(\log (W_j-\openone)-\log (W_j+\openone))
}{(W_j-\openone)^s-(W_j+\openone)^s}
\log\frac{W_j-\openone}{W_j+\openone}\right]
\nonumber.
\end{align}
Finally, calling $V_{12}(s) = V_1(s)+V_2(1-s)$ and $Z_i = Z(G_i)$ we can write 
\begin{align}
	q_s= & -\partial_s \log Q_s \\=& \log\frac{Z_1}{Z_2} + \frac{g_1(s)-g_2(1-s) 
	+\Tr [V_{12}(s)^{-1}V_{12}'(s)]}2 \nonumber  
		\\ &- \frac12\delta^T V_{12}(s)^{-1}V_{12}'(s)V_{12}(s)^{-1} \delta.
\end{align}

\section{Multicopy scaling of the OAQCB}\label{app: multicopy scaling}

Since
\begin{equation}
    \frac{d Q^{N}_{p,(1)}}{dp} = N Q^{N-1}_{p,(1)}\frac{d Q_{p,(1)}}{dp},
\end{equation}
we can write (Eqs.~(\ref{eq: alpha QCB multicopy}) and (\ref{eq: beta QCB multicopy}) in the main text)
\begin{align*}
    &\begin{split}
        \alpha^{(\mathrm{UB},\mathrm{QCB})}_{(N)} &= \exp\left[-N p Q_{p,(1)}^{-1}\frac{dQ_{p,(1)}}{dp}\right] (1-p)Q_{p,(1)}^N\\
        &= \frac{\left(\alpha^{(\mathrm{UB},\mathrm{QCB})}_{(1)}\right)^N}{(1-p)^{N-1}},
    \end{split}\\
    &\begin{split}
        \beta^{(\mathrm{UB},\mathrm{QCB})}_{(N)} &= \exp\left[N(1-p) Q_{p,(1)}^{-1}\frac{dQ_{p,(1)}}{dp}\right] p Q_{p,(1)}^N\\
        &= \frac{\left(\beta^{(\mathrm{UB},\mathrm{QCB})}_{(1)}\right)^N}{p^{N-1}}.
    \end{split}
\end{align*}

We can then use Eqs.~(\ref{eq: alpha QCB multicopy}) and (\ref{eq: beta QCB multicopy}) to recover Eqs.(\ref{eq: gamma alpha}) and (\ref{eq: gamma beta}) from the main text:
\begin{align*}
    &\gamma_{\alpha}^{(\mathrm{UB},\mathrm{QCB})}
    =p Q_{p,(1)}^{-1}\frac{dQ_{p,(1)}}{dp} -\ln \left[Q_{p,(1)}\right],\\
    &\gamma_{\beta}^{(\mathrm{UB},\mathrm{QCB})}
    =-(1-p) Q_{p,(1)}^{-1}\frac{dQ_{p,(1)}}{dp} -\ln \left[Q_{p,(1)}\right].
\end{align*}

\section{Proof that the OAQCB saturates the quantum Hoeffding bound}\label{app: hoeffding bound}

Substituting our expression for $\gamma_{\beta}^{(\mathrm{UB},\mathrm{QCB})}$ into the expression for $b(r,s)$, we get
\begin{equation}
    \begin{split}
        b\left(\gamma_{\beta}^{(\mathrm{UB},\mathrm{QCB})},s\right) =& (1-s)^{-1}\bigg(s(1-p)Q^{-1}_{p,(1)}\frac{dQ_{p,(1)}}{dp}\\
        &+s \ln[Q_{p,(1)}]-\ln[Q_{s,(1)}]\bigg).
    \end{split}\label{eq: b exp}
\end{equation}
To find the maximum achievable decay rate for $\alpha$, subject to the constraint that $\gamma_{\beta}^{(\mathrm{UB},\mathrm{QCB})}\geq r$, we must maximize Eq.~(\ref{eq: b exp}) over $s$ (in the range $0\leq s < 1$). Differentiating with regard to $s$, and noting that $Q_{p,(1)}^{-1}\frac{dQ_{p,(1)}}{dp}=\frac{d}{dp}(\ln[Q_{p,(1)}])$, we get
\begin{equation}
    \begin{split}
        \frac{db\left(\gamma_{\beta}^{(\mathrm{UB},\mathrm{QCB})},s\right)}{ds}
        =&(1-s)^{-2}((1-p)\frac{d}{dp}(\ln[Q_{p,(1)}]) -\\
        &(1-s)\frac{d}{ds}(\ln[Q_{s,(1)}])+\\
        &\ln[Q_{p,(1)}]-\ln[Q_{s,(1)}]).
    \end{split}
\end{equation}
Note that the terms in the numerator are either functions only of $s$ or functions only of $p$. Defining
\begin{equation}
    a_x = (1-x)\frac{d}{dx}(\ln[Q_{x,(1)}])+\ln[Q_{x,(1)}],
\end{equation}
we can write
\begin{equation}
    \frac{db(\gamma_{\beta}^{(\mathrm{UB},\mathrm{QCB})},s)}{ds}=\frac{a_p-a_s}{(1-s)^2}.
\end{equation}

We have a turning point in $b(\gamma_{\beta}^{(\mathrm{UB},\mathrm{QCB})},s)$ iff $a_s=a_p$ (since the denominator is always positive in the range). $s=p$ is therefore a turning point. To determine whether this turning point is a global maximum, we differentiate $a_s$ with regard to $s$. If $\frac{d a_s}{ds}>0$ for $0\leq s <1$, then $s=p$ is a global maximum. If $\frac{d a_s}{ds}\geq 0$ for $0\leq s <1$ (a slightly weaker condition), then the value of $b$ obtained by setting $s=p$ is still $b_{\mathrm{max}}$, even if there is an interval of $s$-values in the neighborhood of $s=p$ that also maximize $b$.
\begin{equation}
    \frac{d a_s}{ds} = (1-s)\frac{d^2}{ds^2}(\ln[Q_{s,(1)}]).
\end{equation}
Since $(1-s)>0$ in our range, the condition for $s=p$ to maximize $b$ reduces to the requirement that the second differential of $\ln[Q_{s,(1)}]$ is positive semidefinite. We therefore need to show that the function $Q_s$ is logarithmically convex (log-convex) in $s$~\cite{kingman_convexity_1961}. This is a stricter condition than convexity, and means that $\ln Q_s$ is also convex (as well as $Q_s$). The second derivative of any convex function is non-negative, so it suffices to show that $Q_s$ is log-convex.

The condition for a function, $f$, to be log-convex is
\begin{equation}
    f(tx_1 + (1-t)x_2)\leq f(x_1)^t f(x_2)^{1-t},
\end{equation}
for $0\leq t \leq 1$~\cite{kingman_convexity_1961}. From Ref.~\cite{kingman_convexity_1961}, we have that the set of log-convex functions (referred to in Ref.~\cite{kingman_convexity_1961} as superconvex functions) is closed under addition. This means that a linear combination of log-convex functions is also log-convex. Ref.~\cite{kingman_convexity_1961} also shows that if every member of a sequence of functions is log-convex, the limit of the supremum of the sequence (limsup) is also log-convex.

First, let us consider the discrete variable case. Recall Eq.~(\ref{eq: Qs discrete decomp}) (repeated here for convenience), which tells us that for discrete variables, we can decompose $Q_s$ as
\begin{equation*}
    Q_s = \sum_i c_i \lambda_i^s \mu_i^{1-s}.
\end{equation*}
Let $f(s)$ be the function $s \mapsto c \lambda^s \mu^{1-s}$, for some positive numbers $\lambda$ and $\mu$. $f(s)$ is log-convex:
\begin{equation}
    \begin{split}
        f(tx_1 + (1-t)x_2) =& c \lambda^{tx_1 + (1-t)x_2} \mu^{1-(tx_1 + (1-t)x_2)}\\
        =& \left(c^t \lambda^{t x_1} \mu^{t (1-x_1)}\right)\\
        &\times\left(c^{1-t} \lambda^{(1-t) x_1} \mu^{(1-t)(1-x_1)}\right)\\
        =&f(x_1)^t f(x_2)^{1-t}.
    \end{split}
\end{equation}
Since $Q_s$ is a sum of such functions, it is also log-convex. This is true even if the sum is unbounded and $i$ takes values up to $\infty$.

Let us extend this result to continuous variable states. We can use the fact that the set of log-convex functions is closed under limsup. Suppose that, for any pair of continuous variable states, we can define a sequence of log-convex approximations to $Q_s$, $Q_s^{(i)}$, so that the supremum of the bounds tends to $Q_s$ in the limit of $i\to\infty$. Then, $Q_s$ must also be log-convex. One subtlety is that the $Q_s^{(i)}$ must all be lower bounds, so that $Q_s$ is the limsup, rather than just the limit.

Recall Eq.~(\ref{eq: Qs continuous decomp}) (repeated here for convenience), which decomposes $Q_s$ as
\begin{equation*}
    Q_s = \int c_x \lambda_x^s \mu_x^{1-s} dx
\end{equation*}
Let us initially assume that $0\leq x < R$, for finite $R$ (i.e. the integral has finite bounds). We can then define our approximations, $Q_s^{(i)}$, as
\begin{equation}
    \begin{split}
        Q_s^{(i)} = \Delta_i \sum_{j=1}^{i} \left(\inf_{(j-1)\Delta_i\leq x < j\Delta_i} c_x \right) \left(\inf_{(j-1)\Delta_i\leq x < j\Delta_i} \rho_x\right)^s\\
        \times\left(\inf_{(j-1)\Delta_i\leq x < j\Delta_i} \sigma_x\right)^{1-s},
    \end{split}
\end{equation}
where $\Delta_i = \frac{R}{i}$. This is a kind of lower Riemann sum, and it is clear that, for any finite number of samples, $i$, $Q_s^{(i)}$ both lower bounds $Q_s$ and is log-convex. Taking the limit as $i\to\infty$, and therefore as $\Delta_i\to 0$, we get $Q_s$. To extend to an infinite domain for $x$, we truncate the function outside the finite domain $0\leq x < R$, take the limit as $i\to\infty$ and then take the limit again as $R\to\infty$.

We can therefore write
\begin{equation}
    b_{\mathrm{max}}(\gamma_{\beta}^{(\mathrm{UB},\mathrm{QCB})})=b(\gamma_{\beta}^{(\mathrm{UB},\mathrm{QCB})},p)=\gamma_{\alpha}^{(\mathrm{UB},\mathrm{QCB})},
\end{equation}
showing that the OAQCB achieves the best possible scaling, according to the quantum Hoeffding bound.

Note that we have assumed that $\gamma_{\beta}^{(\mathrm{UB},\mathrm{QCB})}$ lies in the range $0<\gamma_{\beta}^{(\mathrm{UB},\mathrm{QCB})}<S(\rho_1\|\rho_2)$. We will now show that this always holds, except at the extremal points ($p=0$ and $p=1$), at which the quantum Stein's lemma holds.

\section{Derivation of results showing the OAQCB saturates the quantum Stein's lemma}\label{app: quantum Stein's lemma}

Eqs.~(\ref{eq: p to 0 lim}) and (\ref{eq: p to 1 lim}) in the main text come from applying Eqs.~(\ref{eq: Qs0}) and (\ref{eq: Qs1}) when taking the limits.

We show that $\gamma_{\beta}^{(\mathrm{UB},\mathrm{QCB})}$ is a non-increasing function of $p$ by rewriting Eq.~(\ref{eq: gamma beta}) as
\begin{equation}
    \gamma_{\beta}^{(\mathrm{UB},\mathrm{QCB})} =-(1-p) \frac{d}{dp}(\ln[Q_{p,(1)}]) -\ln \left[Q_{p,(1)}\right]
\end{equation}
and differentiating it, to get
\begin{equation}
    \frac{d \gamma_{\beta}^{(\mathrm{UB},\mathrm{QCB})}}{dp} =-(1-p) \frac{d^2}{dp^2}(\ln[Q_{p,(1)}]).\label{eq: gamma beta diff}
\end{equation}
Since $Q_s$ is log-convex, the right-hand side of Eq.~(\ref{eq: gamma beta diff}) is negative semi-definite, and so $\gamma_{\beta}^{(\mathrm{UB},\mathrm{QCB})}$ is a non-increasing function of $p$.

\section{Error rates for non-adaptive measurement sequences}\label{app: non-adaptive}

\begin{figure}[tb]
\vspace{+0.1cm}
\centering
\includegraphics[width=0.9\linewidth]{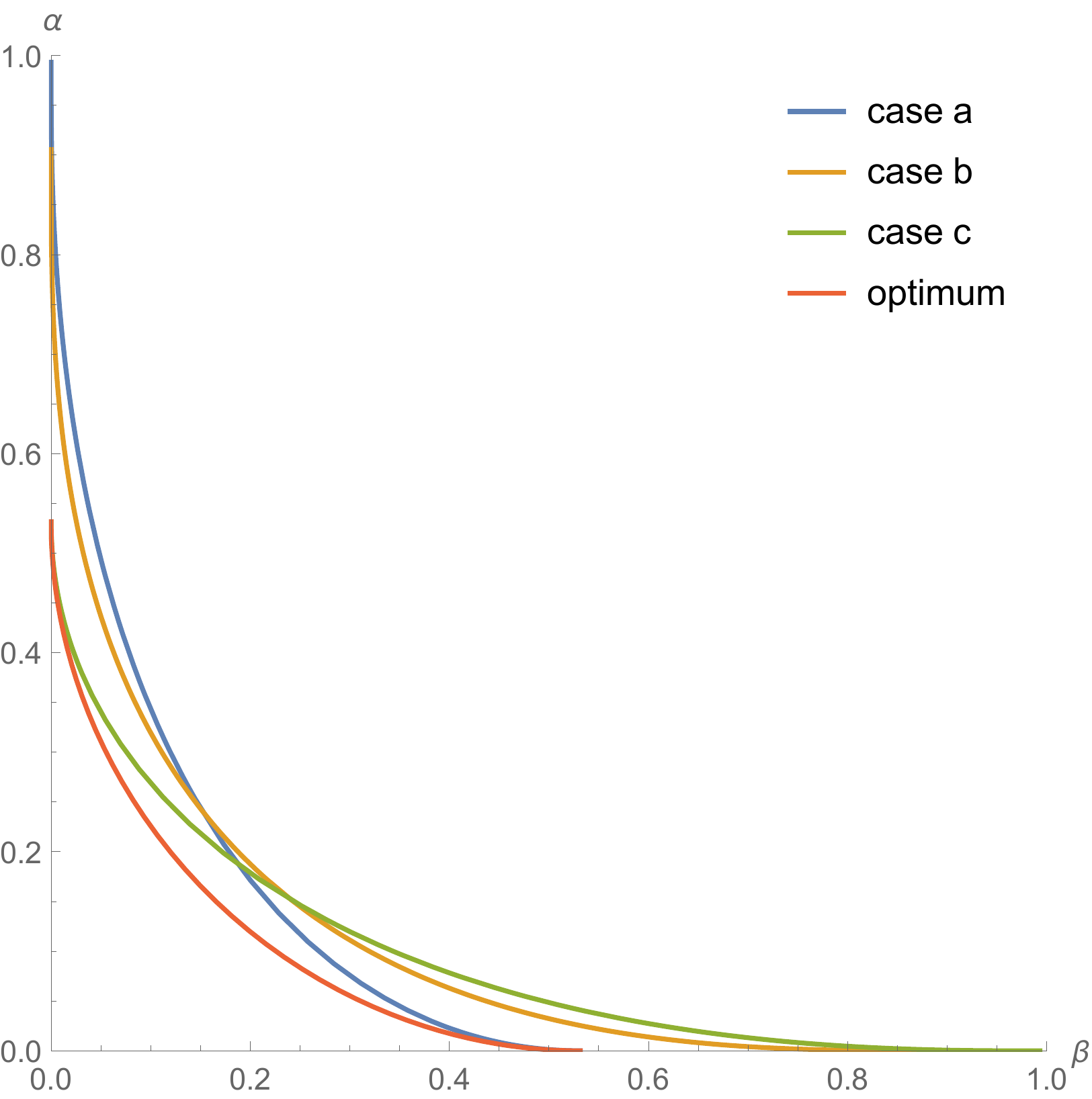}\caption{Type I and II errors for discriminating between a pair of pure, three-copy states, for which the single-copy fidelity is $0.9$. In cases a, b and c, the same (optimal) single-copy measurement is carried out on each subsystem of the state. The cases differ in how we determine the identity of the state from the measurement results. All three methods are worse than the optimal joint measurement, denoted ``optimum" (except at the two points where the optimal curve meets the axes). Case b - where the state is determined using the majority vote - is never better than both cases a and c.}
\label{fig: non-adaptive}
\end{figure}

Consider the measurement sequences described in the main text. The errors for each case can be calculated using Eqs.~(\ref{eq: alpha fid LB param}) and (\ref{eq: beta fid LB param}), and are given by
\begin{align}
    &\alpha^{(\mathrm{a})} = 1 - \left(1-\alpha^{(\mathrm{LB},F_{(1)})}\right)^3,\\
    &\beta^{(\mathrm{a})} = \left(\beta^{(\mathrm{LB},F_{(1)})}\right)^3,
\end{align}
for case a,
\begin{align}
    &\alpha^{(\mathrm{b})} = 3\left(\alpha^{(\mathrm{LB},F_{(1)})}\right)^2-2\left(\alpha^{(\mathrm{LB},F_{(1)})}\right)^3,\\
    &\beta^{(\mathrm{b})} = 3\left(\beta^{(\mathrm{LB},F_{(1)})}\right)^2-2\left(\beta^{(\mathrm{LB},F_{(1)})}\right)^3
\end{align}
for case b, and
\begin{align}
    &\alpha^{(\mathrm{c})} = \left(\alpha^{(\mathrm{LB},F_{(1)})}\right)^3,\\
    &\beta^{(\mathrm{c})} = 1 - \left(1-\beta^{(\mathrm{LB},F_{(1)})}\right)^3.
\end{align}
for case c. All three cases differ from the optimum,
\begin{equation}
    \alpha^{(\mathrm{opt})}=\alpha^{(\mathrm{LB},F_{(3)}=F_{(1)}^3)},
    ~~\beta^{(\mathrm{opt})}=\beta^{(\mathrm{LB},F_{(3)}=F_{(1)}^3)},
\end{equation}
except at the points given by parameter values $p=0$ and $p=1$. This is illustrated in Fig.~\ref{fig: non-adaptive}.

\section{Error rates for adaptive measurement sequences}\label{app: adaptive}

Let us consider the error rates for the adaptive measurement sequence described in the main text (for states with two subsystems).

Recall that we carry out an optimal measurement, with parameter value $p_0$, on the first subsystem. If this measurement tells us that the state is $\rho_{1,1}$, we carry out another optimal measurement with parameter value $p_1$, otherwise we carry out a measurement on the second subsystem with parameter value $p_2$. We use only the result of the second measurement to decide which state we have.

The Type I error for the sequence, $\alpha_{\mathrm{seq}}$, is given by
\begin{equation}
    \begin{split}
        \alpha_{\mathrm{seq}} =& (1-\alpha^{(\mathrm{LB},F_1)}[p_0])\alpha^{(\mathrm{LB},F_2)}[p_1]\\
        &+\alpha^{(\mathrm{LB},F_1)}[p_0]\alpha^{(\mathrm{LB},F_2)}[p_2]
    \end{split}\label{eq: alpha adaptive}
\end{equation}
and the Type II error, $\beta_{\mathrm{seq}}$, is given by
\begin{equation}
    \begin{split}
        \beta_{\mathrm{seq}} =& \beta^{(\mathrm{LB},F_1)}[p_0]\beta^{(\mathrm{LB},F_2)}[p_1]\\
        &+(1-\beta^{(\mathrm{LB},F_1)}[p_0])\beta^{(\mathrm{LB},F_2)}[p_2].
    \end{split}\label{eq: beta adaptive}
\end{equation}

We set (Eq.~(\ref{eq: p1 expression}) in the main text)
\begin{equation*}
    p_1^\mp = \frac{1}{2}\left( 1\mp\sqrt{1-4 p_0(1-p_0)F_1^2}\right).
\end{equation*}
Substituting these values into Eqs.~(\ref{eq: alpha adaptive}) and (\ref{eq: beta adaptive}) (and using Eqs.~(\ref{eq: alpha fid LB param}) and (\ref{eq: beta fid LB param})), we get
\begin{align}
    &\alpha_{\mathrm{seq}} = \frac{2(1-p_0)F_1^2 F_2^2-1+\sqrt{1-4p_0(1-p_0)F_1^2 F_2^2}}{2\sqrt{1-4p_0(1-p_0)F_1^2 F_2^2}},\\
    &\beta_{\mathrm{seq}} = \frac{2p_0 F_1^2 F_2^2-1+\sqrt{1-4p_0(1-p_0)F_1^2 F_2^2}}{2\sqrt{1-4p_0(1-p_0)F_1^2 F_2^2}}.
\end{align}
Now note that
\begin{equation}
    \alpha_{\mathrm{seq}} = \alpha^{(\mathrm{LB},F_1 F_2)}[p_0],~~
    \beta_{\mathrm{seq}} = \beta^{(\mathrm{LB},F_1 F_2)}[p_0],
\end{equation}
so this measurement sequence achieves the optimal errors for discriminating between states with a fidelity of $F_1 F_2$.

\section{Comparison with existing techniques and bounds}\label{app: comparison with existing}

It is possible to calculate the exact ROC using Eq.~(\ref{eq: opt errors}) from the main text. With this is mind, we might ask why we need to use the bounds presented in the main text.

Calculating the eigendecomposition of $(1-p)\rho_2-p\rho_1$ (in order to apply Eq.~(\ref{eq: opt errors})) can be difficult for high-dimensional states and continuous variable (CV) states. Even for Gaussian states (a class of CV states for which calculations are often significantly simpler), the optimal measurement is non-Gaussian, so there is no known simple way to calculate the ROC directly from the first and second moments.

Suppose we only want to numerically find specific points on the ROC. For discrete variable (DV) states, we can formulate the problem of minimizing one type of error with the other fixed as a semidefinite programming problem. Specifically, we can express it as
\begin{alignat*}{2}
    &\mathrm{minimize:} &&1-\Tr [X \rho_2]\\
    &\mathrm{subject~to:}\quad &&X \rho_1 -\alpha\mathcal{I}\geq 0,\\
    & &&X\in\mathcal{H}
\end{alignat*}
where $\mathcal{H}$ is the set of Hermitian matrices with the same dimension as (the matrix representation of) $\rho_{2}$. The optimal $X$ will be the same POVM given by Eq.~(\ref{eq: opt errors}). However, the size of the search space ($\mathcal{H}$) increases with the dimension of the states, so that for high-dimensional states, finding the optimal measurement this way becomes very difficult. For continuous variable states, it is not possible to use this method at all (without some truncation), since they are infinite-dimensional.

Whilst it is also true that calculating the fidelity or $Q_p$, in order to apply the bounds from the main text, also becomes more difficult for higher-dimensional DV states, in many scenarios, these quantities are much easier to calculate. For instance, if the states take tensor-product form (e.g. if we have multiple copies of the same state in both scenarios), we can calculate the fidelity/$Q_p$ on each pair of subsystems individually, and then apply the multiplicativity of fidelity/$Q_p$ over tensor products. For the OAQCB, we can write the following equation for an $N$-partite system (similar to Eqs.~(\ref{eq: alpha QCB multicopy}) and (\ref{eq: beta QCB multicopy}) in the main text):
\begin{align}
    &\alpha^{(\mathrm{UB},\mathrm{QCB})} = \frac{\prod_{i=1}^N \alpha^{(\mathrm{UB},\mathrm{QCB})}_{(i)}}{(1-p)^{N-1}},\\
    &\beta^{(\mathrm{UB},\mathrm{QCB})} = \frac{\prod_{i=1}^N \beta^{(\mathrm{UB},\mathrm{QCB})}_{(i)}}{p^{N-1}},
\end{align}
where $\alpha^{(\mathrm{UB},\mathrm{QCB})}_{(i)}$ and $\beta^{(\mathrm{UB},\mathrm{QCB})}_{(i)}$ comprise the OAQCB calculated for the $i$-th subsystem. In the CV case, if we have Gaussian states, both the fidelity~\cite{banchi2015quantum} and $Q_p$~\cite{pirandola_computable_2008} can be calculated directly from the first and second moments. 

Conversely, the trace norm for tensor product states cannot be calculated using the tensor products on individual subsystems. Similarly, if we want to find the minimum errors numerically, using semidefinite programming, the operators that we minimize over will have the dimension of the entire system. If both states are pure, we can constrain the possible measurements to be convex combinations of measurements in tensor-product form (see Section~\ref{section: sequences} in the main text), but this simplification still does not reduce the problem to minimizing over each subsystem individually. This is clear from the fact that to find a single point on the ROC for discriminating between two pure bipartite systems, one must find two points on the ROC for one of the subsystems, since the optimal measurement can be expressed as an adaptive sequence of measurements on individual subsystems.

If we have a large number of copies of the same state, we can apply the quantum Stein's lemma or the quantum Hoeffding bound to compute the optimal errors. However, these are only tight asymptotically (i.e. for large numbers of copies, $N$), because they govern how quickly the errors exponentially decay. They do not tell us about any sub-exponential terms (such as constant prefactors) that the errors may have. Whilst these terms may become negligible for large $N$, there are situations in which $N$ is small enough that we still care about sub-exponential terms, but still large enough that it is difficult to calculate the exact error values. Recall that the dimension of the states grows exponentially with $N$, so $N$ does not need to be very large for it to be difficult to calculate the ROC exactly.

One task that constitutes an interesting extension to the problem of discriminating between $N$ copies of one of two states is that of discriminating between two different sequences of $N$ states. In such a scenario, it may not be possible to use the quantum Stein's lemma/quantum Hoeffding bound at all. For large $N$, it would also be difficult to calculate the exact errors using the quantum Neyman-Pearson relation. For instance, consider sequences of states whose $n$-th elements are of the form
\begin{equation}
    \rho_{1}^n = \begin{pmatrix}
    \frac{1}{2} &\frac{1}{4n}\\
    \frac{1}{4n} &\frac{1}{2}
    \end{pmatrix},\quad
    \rho_{2}^n = \begin{pmatrix}
    \frac{1}{2} &-\frac{1}{4n}\\
    -\frac{1}{4n} &\frac{1}{2}
    \end{pmatrix}.
\end{equation}
This could model, for instance, a situation in which we can interact with a system multiple times, but each time the interaction is weaker or noisier than the previous. In this scenario, it is not possible to use the quantum Stein's lemma to bound the asymptotic discrimination error. However, we can easily calculate the fidelity and can even take the limit as $N\to\infty$. We can also calculate the OAQCB for any $N$ and for $N\to\infty$.

There are therefore two main scenarios in which the bounds we present are useful. The first is when dealing with CV states (and particularly Gaussian states), since it can be difficult to calculate the exact errors, even in the single-copy case. The second is for tensor-product states that are too large to easily calculate the exact errors for, but for which we still care about the exact values, rather than the asymptotic error exponents (or for which the copies are non-identical).

The hypothesis testing relative entropy~\cite{wang_one-shot_2012}, defined as
\begin{equation}
    D_H^{\epsilon}(\rho\|\sigma)=-\log_2 \inf_{\substack{0\leq Q\leq \mathcal{I},\\ \Tr[Q\rho]\geq 1-\epsilon}}\Tr[Q\sigma],
\end{equation}
is another way of expressing the ROC. By definition,
\begin{equation}
    D_H^{\alpha^*}(\rho_1\|\rho_2)=-\log_2 \beta^*,\quad
    D_H^{\beta^*}(\rho_2\|\rho_1)=-\log_2 \alpha^*.
\end{equation}

From Ref.~\cite{wang_one-shot_2012}, we have the following upper bound on the hypothesis testing relative entropy:
\begin{equation}
    D_H^{\epsilon}(\rho\|\sigma) \leq \frac{S(\rho\|\sigma)+h(\epsilon)}{1-\epsilon},
\end{equation}
where $h$ is the binary entropy function, defined by
\begin{equation}
    h(\epsilon)=-\epsilon\log_2(\epsilon) - (1-\epsilon)\log_2(1-\epsilon),
\end{equation}
and $S$ is the standard quantum relative entropy, but expressed in bits (rather than nats, as was done previously), for consistency with Ref.~\cite{wang_one-shot_2012}. This leads to two different lower bounds on the ROC:
\begin{align}
    &\alpha^* \geq 2^{-\frac{S(\rho_2\|\rho_1)+h(\beta^*)}{1-\beta^*}},\label{eq: htre bound alpha}\\
    &\beta^* \geq 2^{-\frac{S(\rho_1\|\rho_2)+h(\alpha^*)}{1-\alpha^*}}.\label{eq: htre bound beta}
\end{align}
Note that $S(\rho_2\|\rho_1)$ and $S(\rho_1\|\rho_2)$ are generally different, so these bounds are not mirror images of each other.

If the QRE $S(\rho_2\|\rho_1)$ is not infinite (i.e. if the support of $\rho_2$ lies entirely within the support of $\rho_1$), then the bound defined by Eq.~(\ref{eq: htre bound alpha}) runs between the points $(0,2^{-S(\rho_2\|\rho_1)})$ and $(1,0)$ (since this bound is not symmetric, we specify that points are written in the form $(\beta,\alpha)$). Crucially, this means that for any $\beta<1$, $\alpha$ will be non-zero, and so this bound will beat the fidelity-based lower bound over some range of $\beta$ values. If $S(\rho_2\|\rho_1)$ diverges, then the bound is trivial (stating that, for any $\beta^*$, $\alpha^*\geq 0$). Similarly, if $S(\rho_1\|\rho_2)$ does not diverge, the bound defined by Eq.~(\ref{eq: htre bound beta}) runs between the points $(0,1)$ and $(2^{-S(\rho_2\|\rho_1)},0)$, and so this bound will beat the fidelity-based lower bound over some range of $\alpha$ values.

There is no contradiction with the fact that the fidelity-based lower bound is exact for pure states, since $S(\rho_1\|\rho_2)$ and $S(\rho_1\|\rho_2)$ both diverge for pure states ($S(\rho_1\|\rho_2)$ diverges if $\rho_2$ is pure and $S(\rho_2\|\rho_1)$ diverges if $\rho_1$ is pure). In fact, if the support of $\rho_1$ lies within the support of $\rho_2$, it is not possible to set $\alpha=0$ without setting $\beta=1$ (and vice versa, if the support of $\rho_2$ lies within the support of $\rho_1$), so it is not surprising that there exists a tighter bound than the fidelity-based lower bound for small $\alpha$ or $\beta$ in such cases.

The QRE is not a distance metric, and even a small (in terms of trace norm) change in the states $\rho_1$ and $\rho_2$ can result in one or both of the relative entropies varying greatly or even diverging. If the states are not diagonalizable in the same basis and if either of the states is not full rank, at least one of the relative entropies will diverge (and the corresponding bound will become trivial).

Finally, note that $F^2\geq 2^{-S}$, where $S$ is either $S(\rho_1\|\rho_2)$ or $S(\rho_1\|\rho_2)$~\cite{audenaert_comparisons_2014}, so the fidelity-based lower bound will always (except in the $F^2 = 2^{-S}$ case) beat the bounds in Eqs.~(\ref{eq: htre bound alpha}) and (\ref{eq: htre bound beta}) over some range of values (but over a different range for each bound). If both $S(\rho_1\|\rho_2)$ and $S(\rho_2\|\rho_1)$ are close to $F^2$, it may be the case that by choosing whichever is tighter of Eqs.~(\ref{eq: htre bound alpha}) and (\ref{eq: htre bound beta}), we can beat the fidelity-based lower bound over the entire range. However, we find numerically that for states with a sufficiently high fidelity (more than $\sim 0.94$), the fidelity-based bound beats the bounds based on the QRE over some range ($\alpha$ and $\beta$ not close to $0$) even when $F^2 = 2^{-S(\rho_1\|\rho_2)} = 2^{-S(\rho_2\|\rho_1)}$.

\end{document}